\documentclass[twocolumn,epsf,epsfig,showpacs,prd]{revtex4}
\usepackage{graphicx}
\usepackage{dcolumn}
\usepackage{bm}
\usepackage{amsfonts}

\usepackage[dvips]{color}
\definecolor{Black}{named}{Black}
\definecolor{Blue}{named}{Blue}
\definecolor{Red}{named}{Red}

\def\be{\begin{equation}}
\def\ee{\end{equation}}
\def\bea{\begin{eqnarray}}
\def\eea{\end{eqnarray}}

\def\edth{\;\raise1.0pt\hbox{$'$}\hskip-6pt\partial\;}
\def\baredth{\;\overline{\raise1.0pt\hbox{$'$}\hskip-6pt
\partial}\;}

\begin{document}


\title{Rotation of Linear Polarization Plane and Circular Polarization 
from Cosmological Pseudoscalar Fields}
\author{Fabio Finelli$^{a,b,c}$, Matteo Galaverni$^{d,a,c}$}

\affiliation{$^a$INAF-IASF Bologna, 
Via Gobetti 101, I-40129 Bologna - 
Italy}
\affiliation{$^b$INAF/OAB, Osservatorio Astronomico di Bologna,
Via Ranzani 1, I-40127 Bologna - Italy}
\affiliation{$^c$INFN, Sezione di Bologna,
Via Irnerio 46, I-40126 Bologna - Italy}
\affiliation{$^d$Dipartimento di Fisica, Universit\`a di Ferrara,
Via Saragat 1, I-44100 Ferrara - Italy}

\begin{abstract} 
We discuss the rotation of the linear polarization plane 
and the production of circular polarization generated
by a cosmological pseudoscalar field.
We compute analytically and numerically 
the propagation of the Stokes parameters from the 
last scattering surface for an oscillating and a monotonic 
decreasing pseudoscalar field.
For the models studied in this paper, we show the comparison between the
widely
used approximation in which the linear polarization rotation angle is constant in time and the
exact result.

\end{abstract}


\maketitle

\section{Introduction}
In 1977 R. Peccei and H. Quinn \cite{Peccei:1977hh} suggested a solution 
to the strong CP-problem of QCD introducing a new symmetry breaking at 
a given energy scale $f_a$.  The boson associated with this broken global 
symmetry was called {\em axion}. All the physical properties of this  
pseudoscalar field strongly depend on the energy scale $f_a$ at which 
the new symmetry is broken: the particle mass and the coupling constants 
with other particles are inversely proportional to $f_a$. 
Pseudo-Goldstone bosons also arise in many particle physics scenarios 
\cite{kimphysrept}.

Axions and, in general, other pseudoscalar particles 
are among the most favoured particle physics candidates 
for the cold dark matter (CDM) \cite{Kolb:1990vq,Raffelt:1996wa,Sikivie:2006ni}.
They interact with photons according to the lagrangian:
\begin{equation}
\mathcal{L}_{int}=-\frac{g_\phi}{4}\phi  F_{\mu\nu}\tilde{F}^{\mu\nu}\,,
\label{interaction}
\end{equation}
where $g_\phi$ is the coupling constant, 
$F^{\mu\nu}$ is the electromagnetic tensor and 
$\tilde{F}^{\mu\nu}\equiv\frac{1}{2}\epsilon^{\mu\nu\rho\sigma}
F_{\rho\sigma}$ its dual. 
Many constraints on axion derive from this interaction with photons:
laboratory experiments (photon-axion conversion experiments) and astrophysical 
arguments (stellar evolution of red giants) constrain $g_\phi$ to be small.
One of the most stringent experimental bound 
($g_\phi<8.8\times10^{-20}$ $\mbox{eV}^{-1}$ for $m_a<0.02$ eV) 
is obtained by the CAST experiment \cite{Andriamonje:2007ew} constraining the 
axion-photon conversion for solar axions. 
This limit supersedes the one obtained 
from the duration of the helium burning time in horizontal-branch 
stars in globular clusters: $g_\phi\lesssim 10^{-19}$ $\mbox{eV}^{-1}$ 
\cite{Raffelt:1996wa,Raffelt:2006cw}. 

In this paper we wish to study in detail the coupling 
of such a pseudoscalar field with photons. 
The interaction in Eq.~(\ref{interaction}) modifies the polarization 
of an electromagnetic wave propagating  along intervening magnetic fields,
or through a slowly varying background field $\phi$ \cite{Harari:1992ea}.
Here we are interested in the second case, which does not require the 
presence of a magnetic field (note that in the first case 
the polarization is also modified in absence of axions, an effect known as Faraday rotation). 
We consider the time dependent pseudoscalar condensate as dark matter or 
part of it and study the impact of its time derivative on 
the polarization of the photons.
As a consequence of its coupling with a pseudoscalar field, 
the plane of linear polarization of light 
is rotated 
(\textit{cosmological birefringence}) \cite{Carroll:1991zs,Carroll:1997tc}. 

In the case of Cosmic Microwave Background (CMB) photons,
we pay attention to the rotation along the path between the last 
scattering surface (LSS) 
and the observer, modifying the polarization pattern generated by Thomson 
scattering at LSS \cite{Lue:1998mq}. 
This rotation induced by the pseudoscalar interaction 
modifies the gradient and curl of the polarization 
pattern ($E$ and $B$ following \cite{Zaldarriaga:1996xe}), 
creating $B$ modes from $E$ modes.
The parity violating nature of the interaction generates non-zero parity-odd 
correlators ($T \, B$ and $E \, B$)
which would be otherwise vanishing for 
the standard Gaussian cosmological case \cite{Zaldarriaga:1998rg,Hu:1997hv}.  
In particular the $T\,B$ power spectrum may be very useful to constrain the coupling constant $g_\phi$ between photons and pseudoscalars,
since it is larger than the auto and cross power spectra in polarization; in general, these 
non-standard correlators are already constrained by present data sets 
\cite{Feng:2006dp,Cabella:2007br,Komatsu:2008hk}.

We study two representative examples for the dynamics of 
a pseudo-Goldstone field 
behaving as dark matter (see \cite{Liu:2006uh} 
for a pseudoscalar field model of dark energy): 
the oscillating and a monotonic decreasing behavior.
In the latest case we study analytically the problem, 
whereas in the former numerically and analytically. 
The case of a field growing linearly in time has been studied in 
\cite{Balaji:2003sw}.
We compare the polarization power spectra obtained describing the rotation of linear polarization 
with a time dependent angle with the ones obtained considering a constant rotation angle.

Our paper is organized as follows.
We write the relevant equations for 
the electromagnetic gauge potential coupled to a pseudoscalar field in Section
II. We review the Stokes parameters for a monochromatic electromagnetic plane 
wave and 
the Boltzmann equation for CMB photons coupled to pseudoscalars 
in Section III. 
In Section IV we write the Stokes parameters in terms of the 
left and right polarizations gauge potential and solve the 
differential equations for the latter for oscillating behavior 
of the pseudoscalar field. In a similar way Section V is dedicated to the 
monotonic behavior of the pseudoscalar field. 
In Section VI we test the constant rotation angle approximation.
We conclude in Section VII. We work in units where the 
speed of light is equal to one ($c=1$).

\section{Electrodynamics Coupled to a Pseudoscalar Field}
The lagrangian density $\mathcal{L}$ for the photons and 
the pseudoscalar field $\phi$ is \cite{garretson} (following the notation of \cite{finelli_1}): 
\begin{equation}
\mathcal{L}=-\frac{1}{4}F_{\mu\nu} F^{\mu\nu}-\frac{1}{2}\nabla_{\mu}\phi\nabla^{\mu}\phi -V(\phi)-
\frac{g_{\phi}}{4}\phi F_{\mu\nu}\tilde{F}^{\mu\nu}\,.
\end{equation}
The Euler-Lagrange equations resulting from this lagrangian are:
\begin{eqnarray}
\label{eq:phi}
\Box \phi\equiv\nabla_{\mu} \nabla^{\mu} \phi&=& \frac{dV}{d\phi}+ \frac{g_{\phi}}{4} F_{\mu\nu} \tilde{F}^{\mu\nu}\,,\\
\label{eq_F}
\nabla_{\mu}F^{\mu\nu}&=&-g_{\phi}(\nabla_{\mu}\phi)\tilde{F}^{\mu\nu}\,,\\
\nabla_{\mu}\tilde{F}^{\mu\nu}&=&0\;.
\end{eqnarray}
Using the definition of the electromagnetic tensor $F^{\mu\nu}\equiv\nabla^{\mu}A^{\nu}-\nabla^{\nu}A^{\mu}$ 
Eq.~(\ref{eq_F}) becomes:
\begin{equation}
\label{eq_F2}
\Box A_{\nu}-\nabla_{\nu}\left(\nabla_{\mu}A^{\mu}\right)-{R^{\mu}}_{\nu} A_{\mu}=-\frac{g_{\phi}}{2} (\nabla_{\mu} \phi) {{\epsilon^{\mu}}_{\nu}}^{\rho\sigma}F_{\rho\sigma}\;.
\end{equation}
The complete antisymmetric tensor contain the determinant of the metric $g$ 
and $\left[\cdots\right]$ guarantees anti-symmetry in the four indexes \cite{misner}:
\begin{eqnarray}
\epsilon_{\alpha\beta\gamma\delta}&=&\sqrt{-g}\left[\alpha\beta\gamma\delta\right]\,, \\ \epsilon^{\alpha\beta\gamma\delta}&=&-\left(\sqrt{-g}\right)^{-1}\left[\alpha\beta\gamma\delta\right]\,.
\end{eqnarray}
For a spatially flat Friedmann-Robertson-Walker universe the metric is:
\begin{equation}
ds^2=-dt^2+a^2(t) d \boldsymbol{x}^2=a^2(\eta)\left[-d\eta^2+d \boldsymbol{x}^2 \right]\,,
\end{equation}
where $t$ is the cosmic time, $\eta$ is conformal time 
and $\boldsymbol{x}$ denotes the space coordinates. 
We consider a plane wave propagating along $\boldsymbol{\hat{n}}$ 
in Coulomb gauge ($\nabla\cdot\boldsymbol{A}=0$). 
If $\boldsymbol{\hat{n}}$ is aligned with the $z$ axis 
and neglecting the spatial variation of the pseudoscalar 
field $\phi=\phi(\eta)$, the two relevant components of Eq.~(\ref{eq_F2}) are:
\begin{eqnarray}
A_{x}^{\prime\prime}\left(\eta,z\right)-\frac{\partial^2 A_{x}\left(\eta,z\right)}{\partial z^2}&=& g_{\phi} \phi^{\prime} \frac{\partial A_{y}\left(\eta,z\right) }{\partial z}\,,\\
A_{y}^{\prime\prime}\left(\eta,z\right)-\frac{\partial^2 A_{y}\left(\eta,z\right)}{\partial z^2}&=&- g_{\phi} \phi^{\prime} \frac{\partial A_{x}\left(\eta,z\right) }{\partial z}\,.
\end{eqnarray}  
Defining Fourier transform as $\tilde{A}_{x,y}(k,\eta)=(2\pi)^{-1}\int e^{i k z} A_{x,y}(\eta,z) dz$ 
the previous equations become:
\begin{eqnarray}
\tilde{A}_{x}^{\prime\prime}(k,\eta)+k^2 \tilde{A}_{x}(k,\eta)+ g_{\phi} \phi^{\prime} i k \tilde{A}_{y}(k,\eta)=0\,,\\
\tilde{A}_{y}^{\prime\prime}(k,\eta)+k^2 \tilde{A}_{y}(k,\eta)- g_{\phi} \phi^{\prime} i k \tilde{A}_{x}(k,\eta)=0\,.
\end{eqnarray}
where $k$ is the Fourier conjugate of $z$.\\
These equations can be decoupled introducing $\tilde{A}_{\pm}(k,\eta)=\tilde{A}_{x}(k,\eta) \pm i \tilde{A}_{y}(k,\eta)$, left and right components of the electro-magnetic vector potential:
\begin{equation}
\label{eq:A}
\tilde{A}_{\pm}^{\prime\prime}(k,\eta)+\left[ k^2 \pm g_{\phi} \phi^{\prime}  k \right] \tilde{A}_{\pm}(k,\eta)=0\,.
\end{equation}  

\section{Standard Review of Stokes Parameters and Boltzmann Equation}

\subsection{Stokes Parameters}
The complex electric field vector for a plane wave propagating along $\boldsymbol{\hat{z}}$ direction at a point
$(x,y)$ in some transverse plane $z=z_0$ is:
\begin{eqnarray}
\boldsymbol{E}&=&\left(E_x(t)\,,E_y(t)\right)\nonumber\\
&=&\left[ \boldsymbol{\hat{e}_x} \varepsilon_x(t) e^{i \varphi_x(t)}+ \boldsymbol{\hat{e}_y} \varepsilon_y(t) e^{i \varphi_y(t)} \right] e^{-i k t}\,,
\end{eqnarray}
where the physical quantity is the real part of $\boldsymbol{E}$.
For a spatially flat Friedmann-Robertson-Walker metric the relation between the electromagnetic tensor 
and the physical fields is:
\begin{equation}
F_{\mu \nu} = a(\eta)\left( 
\begin{array}{cccc}
  0 & -E_x & -E_y & -E_z\\
  E_x & 0 &  B_z & -  B_y\\
  E_y & -B_z & 0 & B_x\\
  E_z & B_y & -B_x & 0
\end{array}
\right)\,.
\end{equation}
In general we consider quasi-monochromatic waves: the amplitudes ($\varepsilon_x(t)$ and $\varepsilon_y(t)$) 
and the phases ($\varphi_x(t)$ and $\varphi_y(t)$) are slowly varying functions of time respect to the 
inverse frequency of the wave.

The Stokes parameters $I,\, Q,\, U\,\mbox{and}\,V$ are defined as:
\begin{eqnarray}
I&\equiv&\frac{1}{a^2}\left(\left\langle E^*_x(t) E_x(t) \right\rangle + \left\langle E^*_y(t) E_y(t) \right\rangle\right)\,,\\
Q&\equiv&\frac{1}{a^2}\left(\left\langle E^*_x(t) E_x(t) \right\rangle - \left\langle E^*_y(t) E_y(t) \right\rangle\right)\,,\\
U&\equiv&\frac{1}{a^2}\left(\left\langle E^*_x(t) E_y(t) \right\rangle + \left\langle E^*_y(t) E_x(t) \right\rangle\right)\nonumber\\ 
&=& \frac{2}{a^2}\left\langle \varepsilon_x \varepsilon_y \cos \left(\varphi_x-\varphi_y \right)\right\rangle\,,\\
V&\equiv& -\frac{i}{a^2} \left( \left\langle E^*_x(t) E_y(t) \right\rangle - \left\langle E^*_y(t) E_x(t) \right\rangle\right)\nonumber\\
&=& \frac{2}{a^2}\left\langle \varepsilon_x \varepsilon_y \sin \left(\varphi_x-\varphi_y \right) \right\rangle.
\end{eqnarray}
where $\left<\cdots \right>$ denote the {\em ensemble average}, the average over all possible 
realizations of a given quasi-monochromatic wave.
For a pure monochromatic wave ensemble averages can be omitted and the wave is completely polarized:
\begin{equation}
I^2-Q^2-U^2-V^2=0\,.
\end{equation}
The parameter $I$ gives the total intensity of the radiation, $Q$ and $U$ describe linear polarization and $V$ circular polarization. Linear polarization can also be characterized through a vector of modulus: 
\begin{equation}
P_L\equiv\sqrt{Q^2+U^2}\,,
\end{equation}
and an angle $\theta$, defined as: 
\begin{equation}
\theta\equiv\frac{1}{2}\arctan\frac{U}{Q}\,.
\end{equation}

It is important to underline that $I$ and $V$ are physical observables, 
since they are independent on the particular orientation of the 
reference frame in the plane perpendicular 
to the direction of propagation $\boldsymbol{\hat{n}}$, 
while $Q$ and $U$ depend on the orientation of this basis 
\cite{Kosowsky:1994cy}. After a rotation of the reference 
frame of an angle $\theta$ ($R(\theta)$) 
they transform according to:
\begin{equation}
\label{QU_rot}
\begin{array}{ccc}
Q&\stackrel{R(\theta)}{\longrightarrow}& Q\cos (2\theta)+U\sin (2 \theta)\,,\\
U&\stackrel{R(\theta)}{\longrightarrow}& -Q\sin (2\theta)+U\cos (2 \theta)\,.
\end{array}
\end{equation}  
Also linear polarization, like total intensity and circular polarization, can be described through quantities 
independent on the orientation of the reference frame in the plane perpendicular to the 
direction of propagation of the wave. In the context of CMB anisotropies, 
the linear polarization vector field is usually described
in terms of a gradient-like component ({\em E mode}) and of 
a curl-like component ({\em B mode}). 

In a similar way it is possible to describe the 
electric vector field in the $x-y$ plane through a superposition of left and right circular polarized waves defining:
\begin{eqnarray}
\boldsymbol{\hat{e}_+}\equiv \frac{\boldsymbol{\hat{e}_x}+i\boldsymbol{\hat{e}_y}}{\sqrt{2}} \qquad\mbox{and}\qquad \boldsymbol{\hat{e}_-}\equiv \frac{\boldsymbol{\hat{e}_x}-i\boldsymbol{\hat{e}_y}}{\sqrt{2}}\,.
\end{eqnarray}
In this new basis:
\begin{eqnarray}
I&\equiv&\frac{1}{a^2}\left(\left\langle E^*_+(t) E_+(t) \right\rangle + \left\langle E^*_-(t) E_-(t) \right\rangle\right)\,,\\
Q&\equiv&\frac{1}{a^2}\left(\left\langle E^*_+(t) E_-(t) \right\rangle + \left\langle E^*_-(t) E_+(t) \right\rangle\right)\nonumber\\
 &=& \frac{2}{a^2}\left\langle \varepsilon_+ \varepsilon_- \cos \left(\varphi_+ -\varphi_- \right)\right\rangle\,,\\
U&\equiv&-\frac{i}{a^2}\left( \left\langle E^*_+(t) E_-(t) \right\rangle - \left\langle E^*_-(t) E_+(t) \right\rangle\right)\nonumber\\
&=& \frac{2}{a^2}\left\langle \varepsilon_+ \varepsilon_- \sin \left(\varphi_+-\varphi_- \right) \right\rangle\,,\\
V&\equiv& \frac{1}{a^2}\left(\left\langle E^*_+(t) E_+(t) \right\rangle - \left\langle E^*_-(t) E_-(t) \right\rangle\right)\,.
\end{eqnarray}

The relation between the vector potential and the electric field for a wave 
propagating in a charge-free region is:
\begin{equation}
\boldsymbol{E}=-\frac{\partial \boldsymbol{A}}{\partial t}=-\frac{\boldsymbol{A}^{\prime}}{a}\,,
\end{equation}
According to definition given in the previous section the Stokes Parameters in terms of the vector potential are:
\begin{eqnarray}
\label{def:I}
I&=&\frac{1}{a^4}\left(\left<A_+ ^{\prime *} A_+ ^{\prime} \right>+
\left<A_-^{\prime *} A_- ^{\prime}\right>\right)\,,\\
\label{def:Q}
Q&=&\frac{1}{a^4}\left(\left<A_+ ^{\prime *} A_- ^{\prime}\right>+
\left<A_- ^{\prime *} A_+ ^{\prime}\right>\right)\nonumber\\
&=&\frac{2}{a^4}\Re\left(\left<A_+ ^{\prime *} A_- ^{\prime}\right>\right)\,,\\
\label{def:U}
U&=&-\frac{i}{a^4}\left(\left<A_+ ^{\prime *} A_- ^{\prime}\right>-\left<A_- ^{\prime *} A_+ ^{\prime}\right>\right)\nonumber\\
&=&\frac{2}{a^4}\Im\left(\left<A_+ ^{\prime *} A_- ^{\prime}\right>\right)\,,\\
\label{def:V}
V&=&\frac{1}{a^4}\left(\left<A_+ ^{\prime *} A_+ ^{\prime} \right>-\left<A_-^{\prime *} A_- ^{\prime}\right>\right)\,.
\end{eqnarray}

As we shall see in more detail in the following section, the  
coupling to a cosmological pseudoscalar field induce a physical 
time-dependent rotation of the plane of linear polarization along the 
line of sight, described by:
\be
Q^{\prime}=2\theta^{\prime}(\eta) 
U\quad\mbox{and}\quad U^{\prime}=-2\theta^{\prime}(\eta) Q\,,
\ee
whose solution is:
\begin{equation}
\label{QU_evol}
\begin{array}{ccc}
Q &=& Q_i\cos 2\theta + U_i\sin 2 \theta \,,\\
U &=& -Q_i\sin 2\theta + U_i\cos 2 \theta \,.
\end{array}
\end{equation}
where $Q_i \,, U_i$ are the Stokes parameters at initial time which would 
be otherwise unchanged in absence of the interaction with the pseudoscalar 
field.

\subsection{Boltzmann Equation and cosmological birefringence}
\label{sect_Bolz_a}
In the Boltzmann equations for linear polarization of the radiation density 
contrast averaged over momenta contains 
a mixing term: 
\be
2 \theta^\prime =  g \phi^\prime\,,
\ee
due to the pseudoscalar interaction \cite{Lue:1998mq};
the Boltzmann equation for spin-2 functions $Q\pm i U$ is: 
\bea
& &\Delta^\prime_{Q\pm i U} (k,\eta)+i k \mu \Delta_{Q\pm i U} (k,\eta)\nonumber \\
& & = -n_e \sigma_T a(\eta)\left[\Delta_{Q\pm i U} (k,\eta)+\sum_m\sqrt{\frac{6 \pi}{5}} {}_{\pm2}Y_2^m S_P^{(m)}(k,\eta)\right] \nonumber\\
& &\quad \mp i 2\theta^{\prime}(\eta) \Delta_{Q\pm iU}(k,\eta)\,.
\label{qeu}
\eea
where $\mu$ is the cosine of the angle between the CMB photon direction 
and the Fourier wave vector, $n_e$ is the number density of free electrons, 
$\sigma_T$ is the Thomson cross section,
${}_{s}Y_2^m$ are spherical harmonics with spin-weight $s$, 
and $S_P^{(m)}(k,\eta)$ is the source term for generating linear polarization 
reported in \cite{Liu:2001xe} ($m=0,\,\pm1,\,\pm 2$ 
corresponds respectively to scalar, vector, and tenor perturbations):
\be
S_P^{(m)}(k,\eta)=\Delta^{(m)}_{T2}(k,\eta)+12\sqrt{6} \Delta_{+\,,2}^{(m)}(k,\eta)+12\sqrt{6} \Delta_{-\,,2}^{(m)}(k,\eta)\,.
\ee
$\Delta^{(m)}_{Tl}$ and $\Delta^{(m)}_{\pm\,,l}$ are the Fourier transforms 
of the coefficients of the following series:
\bea
& &\Delta_T({\bf x},{\bf \hat{n}},\eta)\nonumber\\
& &\;=\sum_{l\,m} (-i)^{l} \sqrt{4\pi (2l+1)} \Delta^{(m)}_{Tl}({\bf x},\eta) 
Y_l^m({\bf \hat{n}})\,,\\
& &\Delta_{Q\pm i U}({\bf x},{\bf \hat{n}},\eta)\nonumber\\
& &\;=\sum_{l\,m} (-i)^{l} \sqrt{4\pi (2l+1)} \Delta^{(m)}_{\pm\,,\ell}({\bf x},\eta)\, 
{}_{\pm 2} Y_l^m({\bf \hat{n}})\,,
\eea
Note that Eq.~(\ref{qeu}) corrects some typos in Eq.~(1) 
of Ref. \cite{Liu:2006uh}.

The quantity $\Delta_{Q\pm iU}$ is related to the rotation invariant polarization fields 
$\Delta_{E}$ and $\Delta_{B}$ 
through the spin raising ($\edth$) and lowering ($\baredth$) operators:
\begin{eqnarray}
\label{rot_Delta_E}
\Delta_E&\equiv&-\frac{1}{2}\left(\baredth^2 \Delta_{Q+ iU} +\edth^2 \Delta_{Q- iU}\right)\,,\\
\label{rot_Delta_B}
\Delta_B&\equiv&-\frac{i}{2}\left(\baredth^2 \Delta_{Q+ iU} -\edth^2 \Delta_{Q- iU}\right)\,.
\end{eqnarray}

Following the line of sight strategy for scalar perturbations we obtain, in agreement with Ref. \cite{Liu:2006uh}:
\begin{widetext}
\bea
\label{DeltaT}
\Delta_T(k,\eta_0)&=&\int_{\eta_{\rm rec}}^{\eta_0}d\eta\, g(\eta)S_T(k,\eta)j_\ell(k\eta_0-k\eta)\,,\\
\label{DeltaE}
\Delta_E(k,\eta_0)&=&\int_{\eta_{\rm rec}}^{\eta_0}d\eta\, g(\eta)S_P^{(0)}(k,\eta)
\frac{j_\ell(k\eta_0-k\eta)}{\left(k\eta_0-k\eta\right)^2} \cos\left[2 \theta (\eta) \right]\,,\\
\label{DeltaB}
\Delta_B(k,\eta_0)&=&\int_{\eta_{\rm rec}}^{\eta_0}d\eta\, g(\eta)S_P^{(0)}(k,\eta)
\frac{j_\ell(k\eta_0-k\eta)}{\left(k\eta_0-k\eta\right)^2} \sin\left[2 \theta (\eta) \right]\,.
\eea
where $g(\eta)$ is the visibility function, $S_T(k,\eta)$ is the sorce term for temperature anisotropies, ù
and $j_\ell$ is the spherical Bessel function.
The polarization $C_\ell$ auto- and cross-spectra are given by:
\bea
\label{C_ll_EE_dyn}
C_\ell^{EE}&=&\left(4 \pi\right)^2\frac{9\left(\ell+2\right)!}{16\left(\ell-2\right)!}
\int k^2 dk\, \left[\Delta_E(k,\eta_0)\right]^2\,,\\
C_\ell^{BB}&=&\left(4 \pi\right)^2\frac{9\left(\ell+2\right)!}{16\left(\ell-2\right)!}
\int k^2 dk\, \left[\Delta_B(k,\eta_0)\right]^2\,,\\
C_\ell^{EB}&=&\left(4 \pi\right)^2\frac{9\left(\ell+2\right)!}{16\left(\ell-2\right)!}
\int k^2 dk\, \Delta_E(k,\eta_0) \Delta_B(k,\eta_0)\,,\\
C_\ell^{TE}&=&\left(4 \pi\right)^2\sqrt{\frac{9\left(\ell+2\right)!}{16\left(\ell-2\right)!}}
\int k^2 dk\, \Delta_T(k,\eta_0) \Delta_E(k,\eta_0)\,,\\
\label{C_ll_TB_dyn}
C_\ell^{TB}&=&\left(4 \pi\right)^2\sqrt{\frac{9\left(\ell+2\right)!}{16\left(\ell-2\right)!}}
\int k^2 dk\, \Delta_T(k,\eta_0) \Delta_B(k,\eta_0)\,.
\eea
\end{widetext}
In the approximation in which $\theta = \bar{\theta}$, with $\bar{\theta}$ 
constant in time, Eqs.~(\ref{DeltaE}) and (\ref{DeltaB}) simplify 
since $\cos[2 \bar{\theta}] \,, \sin[2 \bar{\theta}]$ can be extracted 
from the integral 
along the line of sight and: 
\bea
\Delta_E^{obs}&=&\Delta_E(\theta=0)\, \cos(2 \bar{\theta})\,,\\
\Delta_B^{obs}&=&\Delta_E(\theta=0)\, \sin(2 \bar{\theta})\,,
\label{Delte}
\eea
and the power spectra are given by \cite{Lue:1998mq,Feng:2006dp}:
\bea
\label{C_ll_EE_constant}
C_\ell^{EE,obs}&=&C_\ell^{EE} \cos^2(2 \bar{\theta})\,,\\
\label{C_ll_BB_constant}
C_\ell^{BB,obs}&=&C_\ell^{EE} \sin^2(2 \bar{\theta})\,,\\
\label{C_ll_EB_constant}
C_\ell^{EB,obs}&=&\frac{1}{2} C_\ell^{EE} \sin(4 \bar{\theta})\,,\\
\label{C_ll_TE_constant}
C_\ell^{TE,obs}&=&C_\ell^{TE} \cos(2 \bar{\theta})\,,\\
\label{C_ll_TB_constant}
C_\ell^{TB,obs}&=&C_\ell^{TE} \sin(2 \bar{\theta})\,.
\eea
The expression for $\bar{\theta}$ to insert in Eqs. (52-58) is:
\be
\bar{\theta} = \frac{g_\phi}{2} \left[ \phi(\eta_0) - \phi(\eta_{\rm rec}) 
\right] \,.
\label{assumption}
\ee
Several limits on the constant rotation angle $\bar{\theta}$ have been 
already obtained 
using current observation of CMBP (see Tab.~\ref{tab:Theta_tab}).

\begin{table}[htbp]
\begin{center}
 \begin{tabular}{|l|c|}
  \hline
     Data set & $\bar{\theta}\;(2\sigma)\;\mbox{[deg]}$\\
    \hline
     WMAP3 and Boomerang (B03) \cite{Feng:2006dp} & $-13.7<\bar{\theta}<1.9$\\
    \hline
     WMAP3 \cite{Cabella:2007br} & $-8.5<\bar{\theta}<3.5$\\
    \hline
     WMAP5 \cite{Komatsu:2008hk} & $-5.9<\bar{\theta}<2.4$\\
    \hline 
     QUaD \cite{Wu:2008qb} & $-1.2<\bar{\theta}<3.9$\\
    \hline   \end{tabular}
\end{center} 
	\caption{Constraints on linear polarization rotation $\bar{\theta}$ in the constant angle approximation.
	}
\label{tab:Theta_tab}
\end{table}

This time independent rotation angle approximation is an operative 
approximation which allows to write Eqs. (\ref{Delte}),
is clearly inconsistent since for $\theta=\mbox{const}$ the term proportional 
to $\theta^\prime$ in the Boltzmann equation (\ref{qeu}) vanishes and therefore 
there is no rotation of 
the linear polarization plane.
See Figs. \ref{plot::EE2_4} and \ref{Bolt_mono_2} for a comparison of this 
approximation 
with a full Boltzmann description of the birefringence effect
for a dynamical pseudoscalar field.

\section{Cosine-type potential}
In this section we assume that dark matter is given by 
massive axions,
$\phi$ is governed by the potential \cite{Kolb:1990vq}:
\be
\label{alps_pot}
V(\phi)= m^2 \frac{f_a^2}{N^2} \left(1-\cos\frac{\phi N}{ f_a}\right)\,,
\ee
where $N$ is the color anomaly of the Peccei-Quinn symmetry.
Here we are interested in the regime where the axion field oscillates 
near the minimum of 
the potential (for simplicity we shall consider $N=1$ in the following): 
$\phi/f_a\ll 1$ and the potential can be approximated with  
$V(\phi) \simeq m^2 \phi^2/2$. In this case $\phi(t)$ satisfies the equation:
\begin{equation}
\ddot{\phi}+3 H \dot{\phi}+m^2 \phi=0\,.
\end{equation}
When $m>3 H$ the scalar field begins to oscillate, and the solution in a matter dominated universe 
($\dot{a}/a=2/3 t$) is \cite{Dine:1982ah}: 
\begin{eqnarray}
& &\phi(t)=t^{-1/2}\left[c_1 J_{1/2}(m t)+c_2 J_{-1/2}(m t) \right]\nonumber\\
& & \stackrel{mt\gg 1}{\simeq} \frac{\phi_0}{mt} \sin(m t) \,,
\end{eqnarray}
where the time-independent coefficients of the Bessel functions $c_1 \,, c_2$ 
depend on the initial conditions.

The averaged energy and pressure densities associated with the field are:
\begin{eqnarray}
\overline{\rho_\phi}&=&\frac{\overline{\dot{\phi}^2}}{2}+\frac{1}{2}m^2 \overline{\phi}^2
\stackrel{mt\gg 1}{\simeq}  \frac{\phi_0^2}{2 t^2}\left[1+\mathcal{O}\left(\frac{1}{mt}\right)^2\right]\,, 
\\
\overline{P_\phi}&=&\frac{\overline{\dot{\phi}^2}}{2}-\frac{1}{2}m^2 \overline{\phi}^2
\stackrel{mt\gg 1}{\simeq}  \frac{\phi_0^2}{2 t^2} \times \mathcal{O}\left(\frac{1}{mt}\right)^2\,,
\end{eqnarray}
where  $\bar{ }$  denotes the average over an oscillation period of the axion condensate. 
Note that we are implicitly assuming that the pseudoscalar field is homogeneous. 
In the context of axion physics, this means that in our observable universe we have just one value for 
the misalignment angle, which means that the PQ symmetry has occurred before or during inflation.

We fix the constant $\phi_0$ comparing $\rho_\phi$ with the energy density in a matter dominated universe:
\begin{equation}
\rho_M=\frac{3 H^2 M_\mathrm{ pl}^2}{8 \pi}=\frac{M_\mathrm{ pl}^2}{6 \pi t^2}
\quad \Longrightarrow \quad \phi_0=\frac{M_\mathrm{ pl}}{\sqrt{3 \pi}}\,,
\end{equation}
\be
\label{phiT1}
\phi(t)\simeq\frac{M_\mathrm{ pl}}{\sqrt{3 \pi} m t}\sin(m t)\,,
\ee
where $M_\mathrm{ pl}\simeq 1.22\times10^{19}$ GeV is the Planck mass.

\begin{figure}
\includegraphics[width=8.6cm]{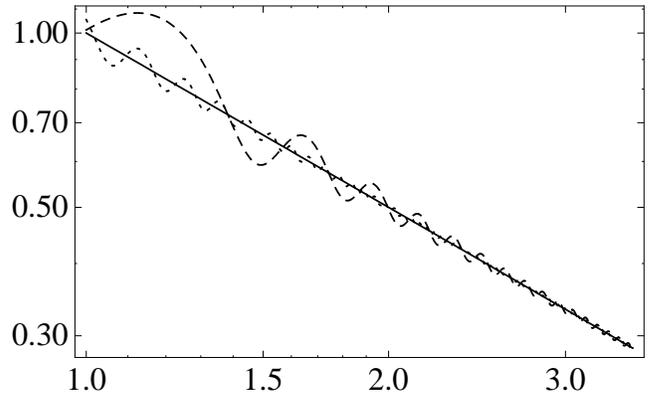}
\caption{Evolution of $\mathcal{H}/\mathcal{H}_{\rm rec}$ in function of conformal time 
for $m=10^{-28}$eV (dashed line), $m=5\times10^{-27}$eV (dotted line) 
and for a matter dominated universe (continuous line), 
from recombination ($\eta_ {\rm rec}$) to $3.5 \eta_{\rm rec}$.
Present time corresponds to $\eta_0=\eta_{\rm rec}\sqrt{1+z_{\rm rec}}\simeq33.18 \eta_{\rm rec}$.
}
\label{fig_H}
\end{figure}

Using the relation between cosmic and conformal time in a universe of matter:
\be
t=\frac{\eta_0}{3}\left(\frac{\eta}{\eta_0}\right)^3\,,
\ee
we find the following approximation for $\phi(\eta)$:
\begin{equation}
\phi(\eta)\simeq\sqrt{\frac{3}{\pi}} \frac{M_\mathrm{ pl}}{m \eta_0 \left(\frac{\eta}{\eta_0}\right)^3}
\sin\left[ m \frac{\eta_0}{3}\left(\frac{\eta}{\eta_0}\right)^3\right] \,,
\end{equation}
and
\begin{eqnarray}
\phi^\prime(\eta) \simeq\sqrt{\frac{3}{\pi}} \frac{M_{\rm pl}}{\eta}& & \left\{
 \cos\left[ m \frac{\eta_0}{3}\left(\frac{\eta}{\eta_0}\right)^3\right] \right. \nonumber\\
 \label{phip_1}
 & & \left.  - \frac{3 \eta_0^2}{m \eta^3}\sin \left[ m \frac{\eta_0}{3}\left(\frac{\eta}{\eta_0}\right)^3\right]\right\}\,.
\end{eqnarray}
If $m$ is not too small the value of $\mathcal{H}\equiv a^\prime / a$ 
obtained with the scalar field density in the Friedmann equation coincides 
with that of a matter dominated universe $\mathcal{H}=2/\eta$ 
once the average through oscillations is performed \cite{Turner:1983he} (see Fig.~\ref{fig_H}).
The derivative can be replaced in Eq.~(\ref{eq:A}) 
for the evolution of the gauge potential: 
\begin{equation}
\label{eq:A_2}
\tilde{A}_{\pm}^{\prime\prime}(k,\eta)+k^2\left[1\pm \Delta(\eta;g_\phi,m,k,\eta_0)  \right] \tilde{A}_{\pm}(k,\eta)=0\,,
\end{equation}
defined the function:
\begin{eqnarray}
\Delta (\eta;g_\phi,m,k,\eta_0)&\equiv& \sqrt{\frac{3}{\pi}}\frac{g_\phi M_{\rm pl}}{k\eta} \left\{\cos\left[ m \frac{\eta_0}{3}\left(\frac{\eta}{\eta_0}\right)^3\right]\right. \nonumber\\
& &\left. -\frac{3 \eta_0^2}{m \eta^3} \sin\left[ m \frac{\eta_0}{3}\left(\frac{\eta}{\eta_0}\right)^3\right] \right\}\,.
\end{eqnarray}
This term, induced by axion-photon coupling, oscillates with frequency proportional 
to the mass of the axion and its amplitude decreases with time.

In the next two subsections we study analytically and numerically Eq.~(\ref{eq:A_2}) 
for different values of the parameters $m$ and $g_\phi$;
we exclude the region where the mass of the pseudoscalar field is so small that the field starts to oscillate 
after equivalence ($m<3 H_{\rm eq}$), 
and the region corresponding to a PQ symmetry broken at energies higher than Planck scale ($f_a>M_{\rm pl}$): 
see Fig.~\ref{mg01}. 

\subsection{Adiabatic solution}
Adiabatic solutions of Eq.~(\ref{eq:A_2}) are:
\begin{equation}
\label{A_adiabatic}
\tilde{A}_s=\frac{1}{\sqrt{2\omega_s}}e^{\pm i \int \omega_s d\eta}\,,
\end{equation}
where $\omega_s(\eta)=k\sqrt{1\pm\frac{g_\phi \phi^{\prime}(\eta)}{k}}=k\sqrt{1\pm\Delta(\eta)}\,$ and $s=\pm\,.$\\
The second derivative respect to conformal time is:
\begin{equation}
\tilde{A}_s^{\prime\prime}=\tilde{A}_s\left(-\omega_s^2+\frac{3 \omega_s^{\prime 2}}{4 \omega_s^2}-\frac{\omega_s^{\prime\prime}}{2 \omega_s^3}\right)\,.
\end{equation}
The adiabatic solution~(\ref{A_adiabatic}) is a good approximation for the vector potential when the terms 
$\frac{3 \omega_s^{\prime 2}}{4 \omega_s^2}$ and $\frac{\omega_s^{\prime\prime}}{2 \omega_s^3}$ are small
compared to $\omega^2_s$:
\begin{eqnarray}
& &\frac{3 \omega_s^{\prime 2}}{4 \omega_s^4}=\frac{3\Delta^{\prime 2}}{16 k^2 \left(1\pm\Delta\right)^3} \ll 1\,,\\
& &\frac{\omega_s^{\prime \prime}}{2 \omega_s^3}=\frac{\pm 2(1\pm \Delta) \Delta^{\prime\prime}-\Delta^{\prime 2}}{8 k^2 \left(1\pm\Delta\right)^3} \ll 1\,.
\end{eqnarray}
If both condition are satisfied and $\Delta\ll 1$:
\begin{eqnarray}
\tilde{A}_\pm&\simeq& \frac{1}{\sqrt{2k \left(1\pm \Delta/4 \right)}} \exp\left[\pm i k\left(\eta \pm \frac{1}{2} \int \Delta d\eta  \right)\right]\nonumber\\
&=& \frac{1}{\sqrt{2k\left(1\pm\pi g_\phi \phi^\prime k\right)}}\exp \left[\pm i\left(k\eta\pm2 \pi g \phi \right) \right]\,.
\end{eqnarray}
In the adiabatic regime the coupling between photons and axions produces a frequency 
independent shift between the two polarized waves, which corresponds  
to a rotation of the plane of linear polarization:
\begin{equation}
\theta_{\rm adiabatic}=\frac{g_\phi}{2} 
\left[\phi(\eta_{0})-\phi(\eta_{\rm rec})\right]\,.
\label{adiabatic}
\end{equation}
This result agrees with the one obtained in Ref. \cite{Harari:1992ea}, 
which therefore holds in the adiabatic regime. 
More important than this, $\theta_{\rm adiabatic} = {\bar \theta}$, i.e. 
Eq. (\ref{adiabatic}) agree with the rotation angle which is approximated 
by Eq. (\ref{assumption}) in the Boltzmann section III.B. 
This agreement is not a coincidence and 
shows the usefulness of studying the gauge potential as done 
in this section: the estimate based on the adiabatic approximation of the 
rotation angle due to cosmological 
birefringence can be also obtained by studying the gauge potential 
$A_s$.

Typically $\phi(\eta_{\rm rec})\gg \phi(\eta_{0})$; from last scattering to 
now $\overline{\rho}\simeq m^2 \overline{\phi}^2 $ so, in a matter dominated universe:
\begin{equation}
\overline{\phi}(\eta)\simeq\sqrt{\frac{3}{8 \pi}}\frac{ M_{\rm pl}\mathcal{H}(\eta) }{m }
\simeq\sqrt{\frac{3}{2 \pi}} \frac{ M_{\rm pl}}{m \eta_0}\left(\frac{\eta_0}{\eta}\right)^3\,.
\end{equation}
An estimate of the angle $\theta_{\rm adiabatic}$ is:
\begin{equation}
\theta_{\rm adiabatic}\simeq g_\phi \sqrt{\frac{3}{8\pi}} \frac{ M_{\rm pl}}{m \eta_0}\left[\left(1+z_{\rm rec}\right)^{3/2}-1\right]\,.
\end{equation}
Note the dependence of $\theta_{\rm adiabatic}$ on the coupling constant and 
on the mass of the pseudoscalar field:
for fixed $g_\phi$ the effect is larger for smaller masses. 

The amplitude of the electromagnetic field changes according to:
\begin{equation}
\left|\boldsymbol{\tilde{E}}\right|^2=\frac{\left|\boldsymbol{\tilde{A}^\prime}\right|^2}{a^2}\simeq
\frac{\omega_s}{2 a^2}\,,
\end{equation}
so the degree of circular polarization evolves according \cite{footnote1,Lee:1999ae}:
\begin{eqnarray}
\tilde{\Pi}_C&=&\frac{\left|\tilde{A}_+^\prime\right|^2 -
\left|\tilde{A}_-^\prime\right|^2}{\left|\tilde{A}_+^\prime\right|^2 +\left|\tilde{A}_-^\prime\right|^2}\nonumber\\
&=&\frac{\sqrt{1+\Delta}-\sqrt{1-\Delta}}{\sqrt{1+\Delta}+\sqrt{1-\Delta}}\simeq\frac{\Delta}{2}=\frac{2 \pi g_\phi \phi^\prime} {k}\,.
\label{circulardegree}
\end{eqnarray}

\subsection{CMBP constraints on the $(m,g_{\phi})$ plane} 
In a flat universe dominated by  dust ($w=0$) plus a component with $w=-1$ 
(cosmological constant) 
the evolution of the scale factor in terms of cosmic time is \cite{Gruppuso:2005xy}:
\be
a(t)=\left(\frac{\Omega_\mathrm{ MAT}}{1-\Omega_\mathrm{ MAT}}\right)^\frac{1}{3}
\sinh \left[ \frac{3}{2}\sqrt{1-\Omega_\mathrm{ MAT}} H_0 t\right]^\frac{2}{3}\,,
\ee
where $\Omega_\mathrm{ MAT}$ is the density parameter for matter nowadays.
The Hubble parameter is:
\be
H=H_0 \sqrt{1-\Omega_\mathrm{ MAT}}\coth\left(\frac{3}{2}\sqrt{1-\Omega_\mathrm{ MAT}}H_0 t\right)\,.
\ee
The pseudoscalar field evolves according to:
\bea
\phi(t)&\stackrel{m t\gg 1}{\simeq}&
 \frac{\phi_0}{\left[\sinh\left(\frac{3}{2}\sqrt{1-\Omega_\mathrm{ MAT}}H_0 t\right) \right]} \nonumber\\
& & \times 
\sin\left[m t \sqrt{1-\left(1-\Omega_\mathrm{ MAT}\right)\left(\frac{3H_0}{2m}\right)^2}\right]\,.
\eea
The energy density is:
\bea
\rho_\phi&=&\frac{\overline{\dot{\phi}^2}}{2}+\frac{1}{2}m^2 \overline{\phi}^2\nonumber\\
&\stackrel{mt\gg 1}{\simeq}&  \frac{m^2\phi_0^2}{2\left[\sinh\left(\frac{3}{2}\sqrt{1-\Omega_\mathrm{ MAT}}H_0 t\right) \right]^{2}}\propto a^{-3}\,.
\eea
Assuming that the axion-like particles contribute to the cold dark matter 
density $\rho_{\phi,\,0}=\Omega_\mathrm{ MAT}\,\rho_\mathrm{ CR,\,0}$ 
(where $\rho_\mathrm{ CR,\,0}$ is the critical density)  
we can estimate $\phi_0$:
\be
\phi_0=\sqrt{\frac{3(1-\Omega_\mathrm{ MAT})}{\pi}}\frac{H_0 M_\mathrm{ pl}}{2  m}\,.
\ee
Therefore the evolution of the pseudoscalar field as a function of cosmic time is:
\bea
\phi(t)
&=&\sqrt{\frac{3\Omega_\mathrm{ MAT}}{\pi}}\frac{H_0 M_\mathrm{ pl}}{2 m a^{3/2}(t)}\nonumber\\
\label{E:phi_t_1}
& &\times
\sin\left[m t \sqrt{1-\left(1-\Omega_\mathrm{ MAT}\right)\left(\frac{3H_0}{2m}\right)^2}\right]\,.
\eea
Note how this equations reduces to Eq.~(\ref{phiT1}) in a matter dominated universe:
$\Omega_\mathrm{ MAT}=1$, $H_0/(2 a^{3/2})=1/(3 t)$.
The linear polarization plane, from last scattering surface, 
rotates according to:
\be
\label{Eq.:rot_anfle}
\theta(t)=\frac{ g_\phi}{2}\left[\phi(t)-\phi(t_\mathrm{rec})\right]\,.
\ee
The Boltzmann equation contains the derivative of the rotation angle respect to of conformal time
(cfr. Eq.~(\ref{qeu})),
so we need the relation between cosmic and conformal time. 
For a particular model with $\Omega_\mathrm{ MAT}=0.3$ it is possible to fit numerically the relation between 
cosmic and conformal time from last scattering to nowadays:
\be
t\simeq \frac{\eta_0}{3.45}\left(\frac{\eta}{\eta_0}\right)^{3.09}\,.
\ee
Replacing this expression in Eq.~(\ref{E:phi_t_1}) we obtain
the evolution of the pseudoscalar field as a function of conformal 
time $\phi=\phi(\eta)$.

\begin{figure}
\begin{center}
\includegraphics[width=8.6cm]{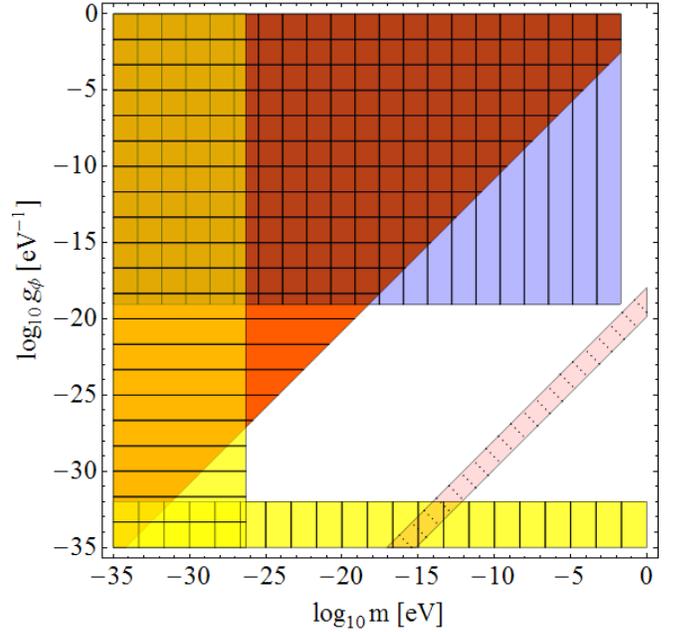}
\caption{Plane $(\log_{10} m\,\left[\mbox{eV}\right],\, \log_{10} g_\phi\, \left[\mbox{eV}^{-1}\right])$:
region excluded by CAST \cite{Andriamonje:2007ew} (blue with vertical lines), 
region where $\left|\theta_A(\Omega_\mathrm{ MAT}=0.3,m,g_\phi)\right|>10$~deg 
(red region with horizontal lines), $\left(m,\,g_\phi\right)$ values expected in main QCD 
axion models (red with dots),
region where the mass of the pseudoscalar field is too small in order
to explain dark matter ($m<3H_\mathrm{ eq}$) (yellow with horizontal lines), 
and
region where PQ symmetry is broken at energies higher than Planck scale 
($f_a>M_\mathrm{ pl}$) (yellow with vertical lines).} 
\label{mg01}
\end{center}
\end{figure}


\begin{figure*}
\begin{center}
\begin{tabular}{c}
\includegraphics[width=8.6cm]{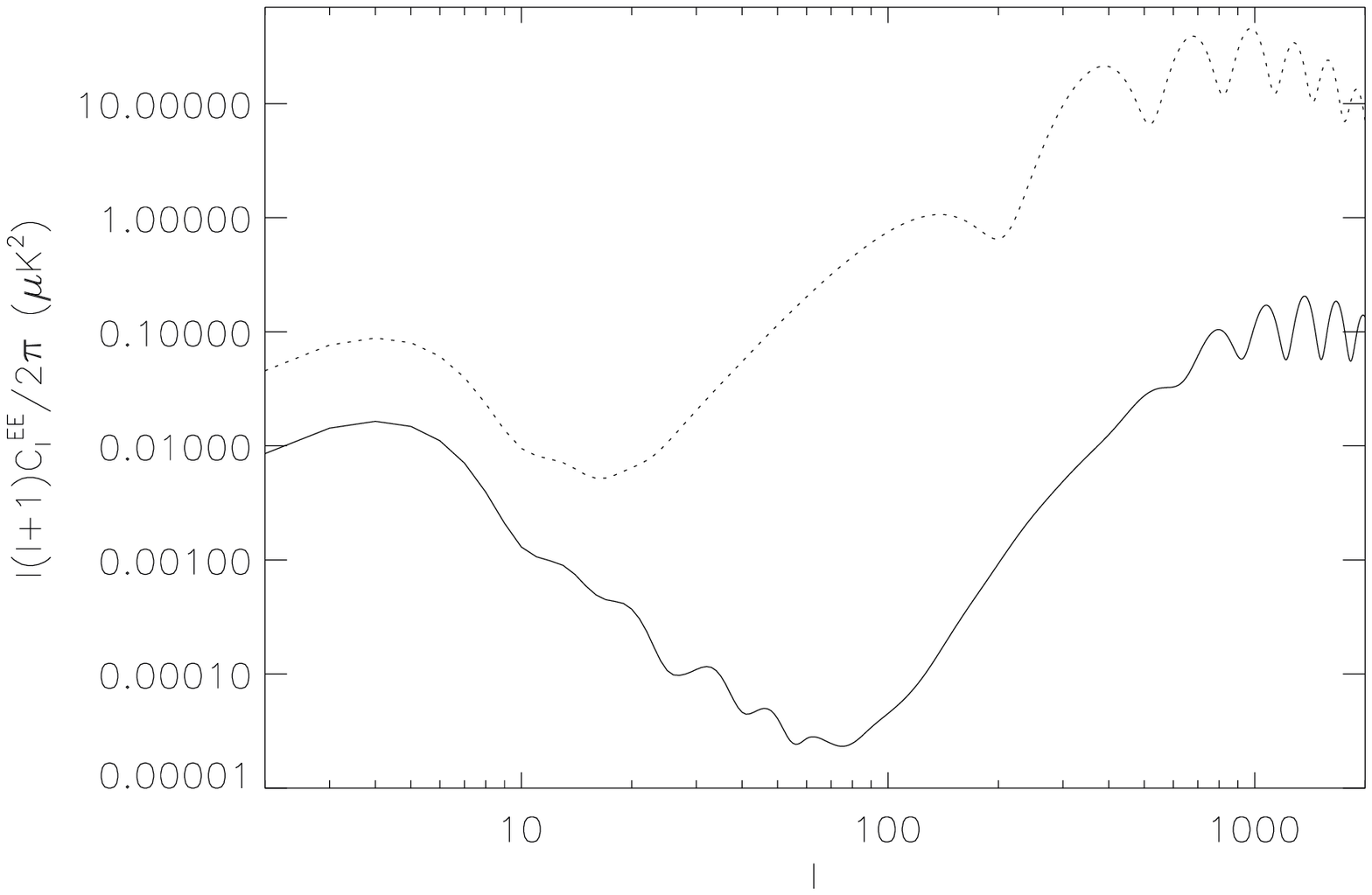}
\includegraphics[width=8.6cm]{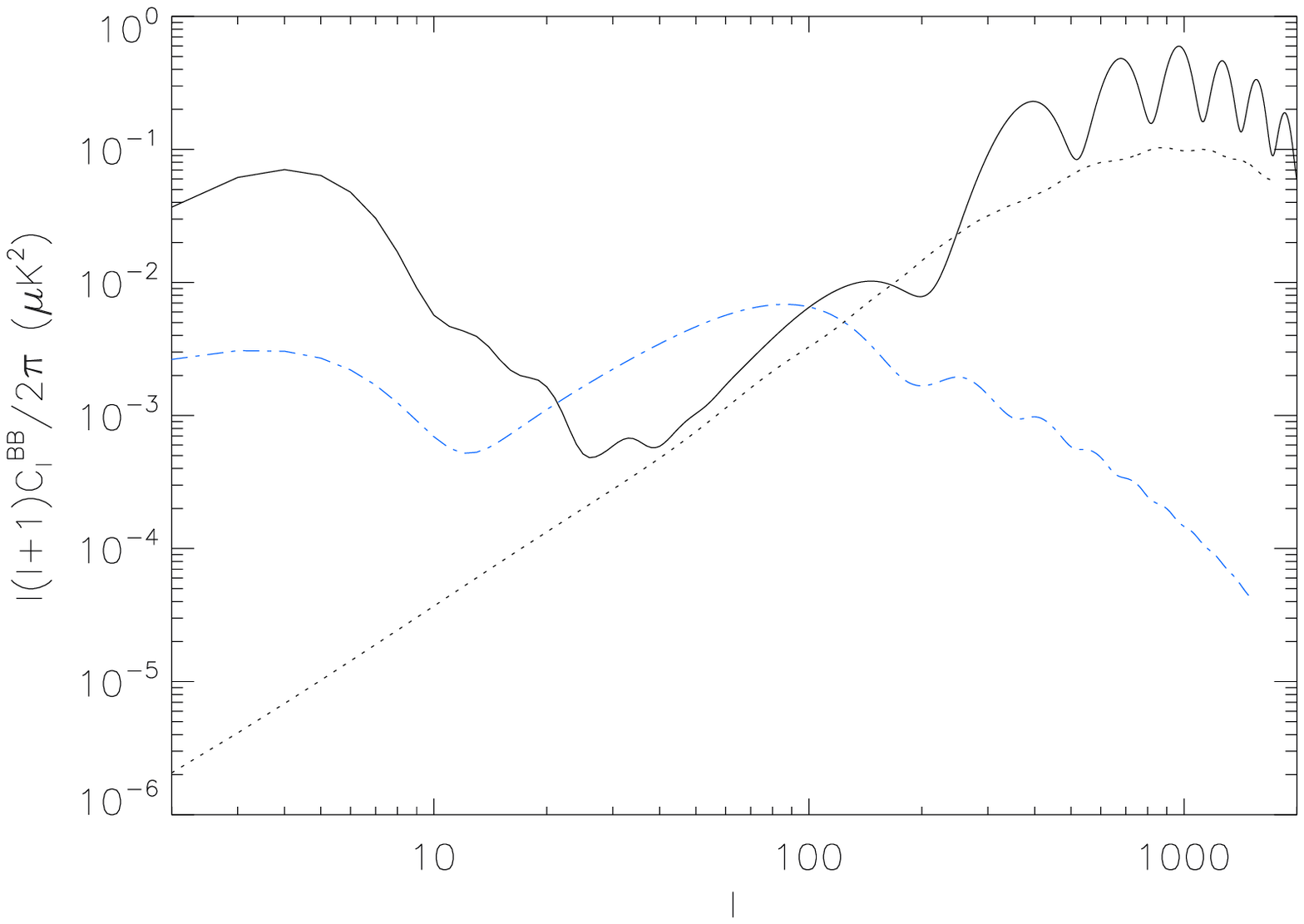}\\
\includegraphics[width=8.6cm]{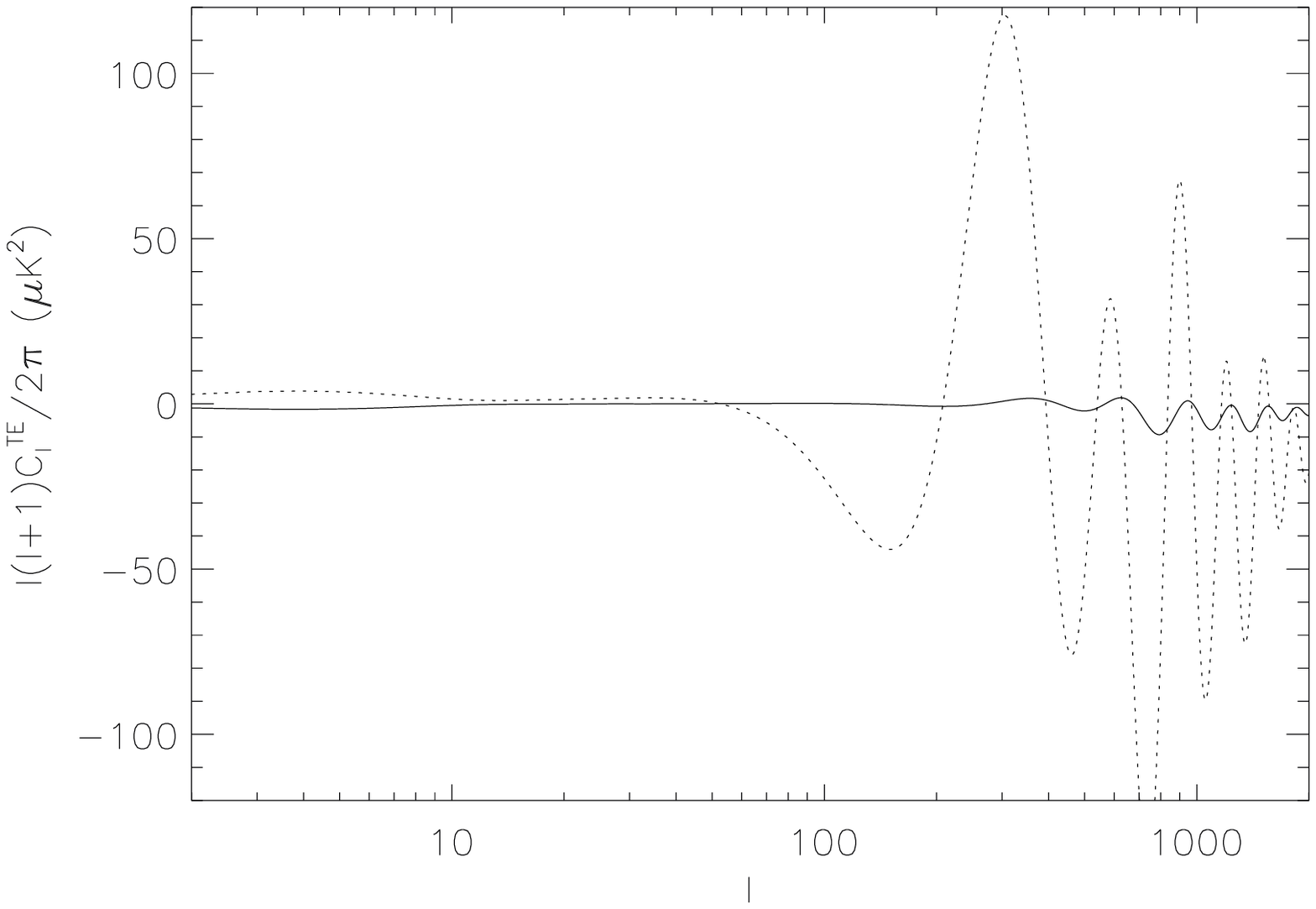}
\includegraphics[width=8.6cm]{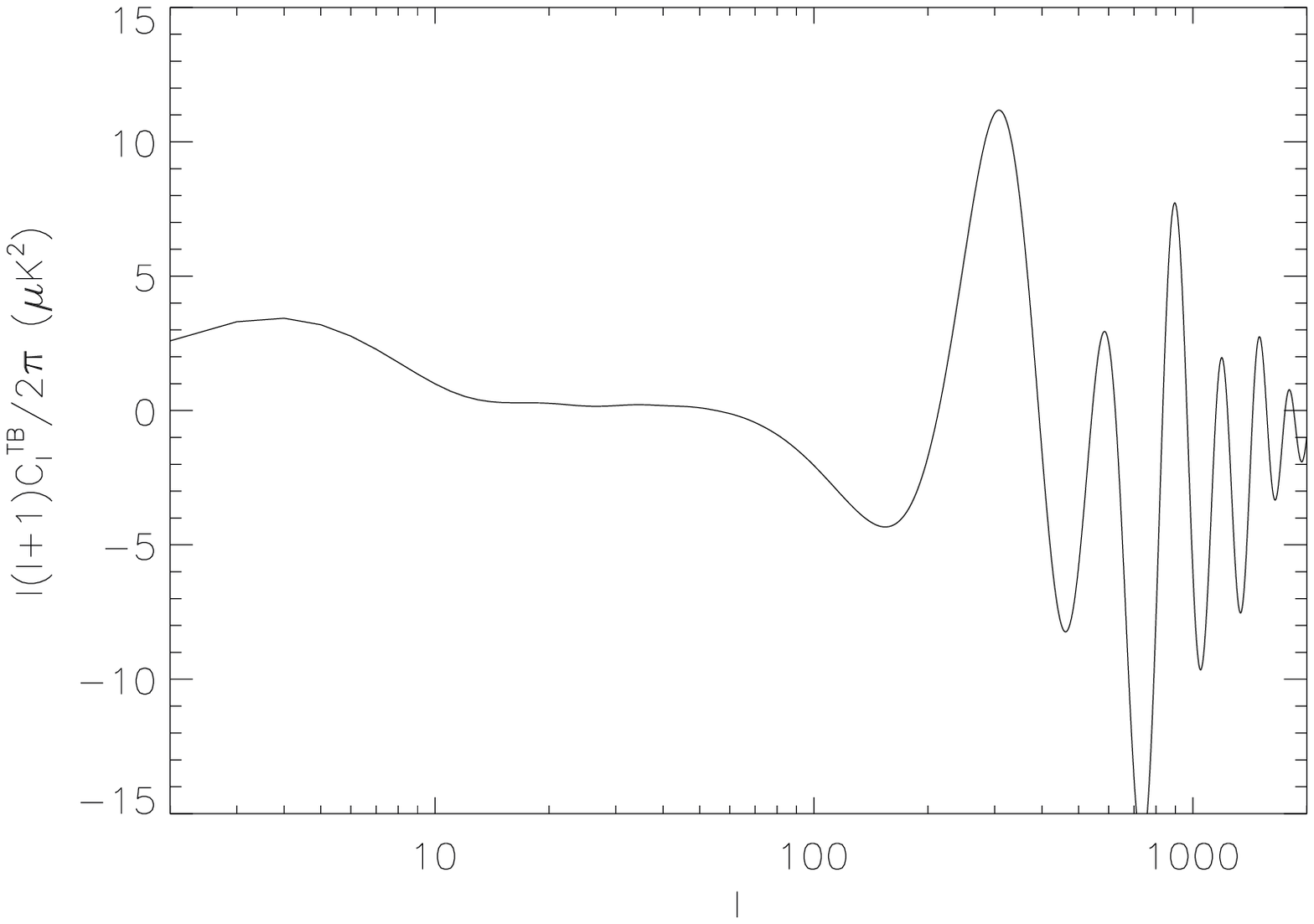}\\
\includegraphics[width=8.6cm]{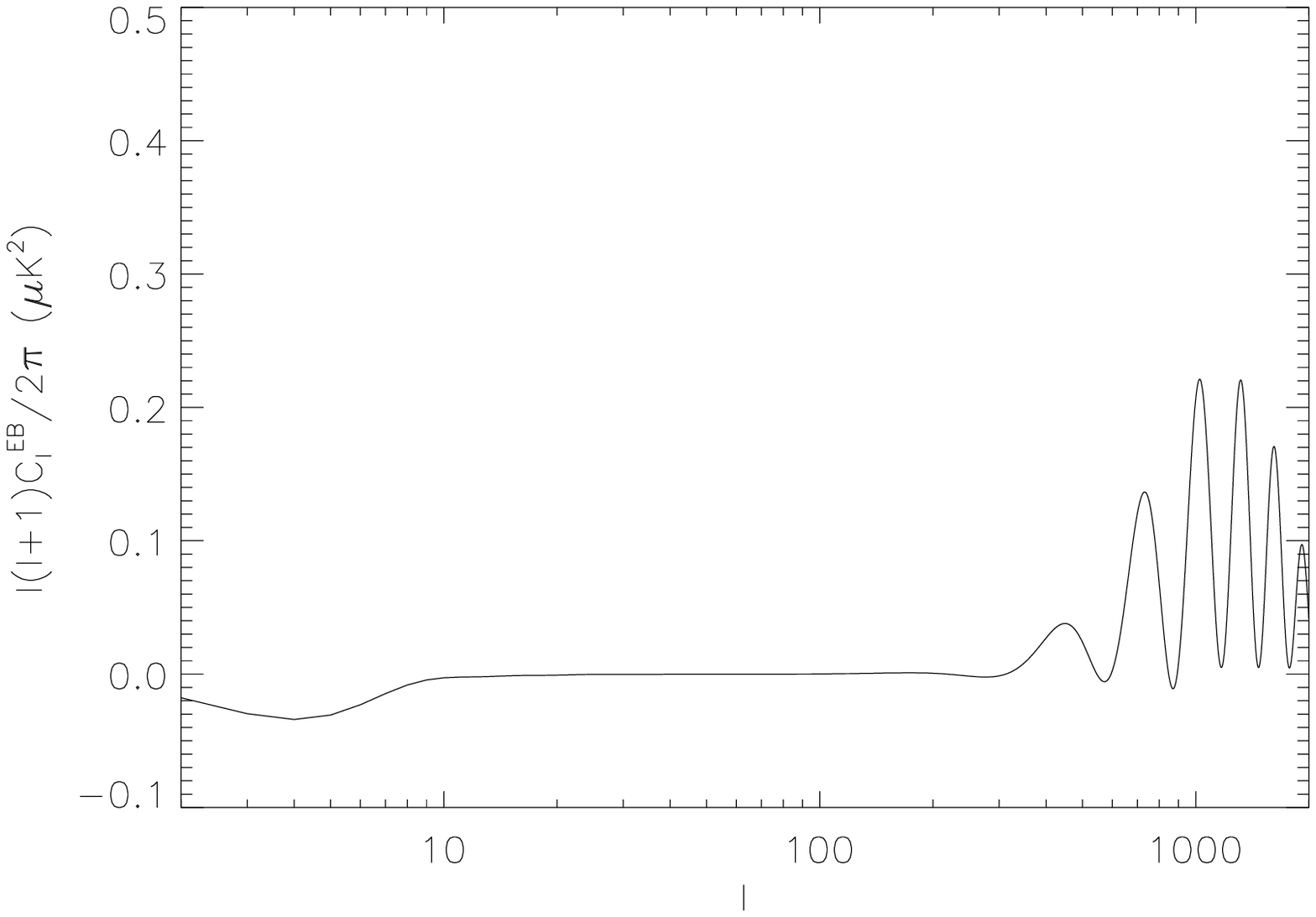}
\end{tabular}
\caption{$E\,E$ (a), $B\,B$ (b), $T\,E$ (c), $T\,B$ (d), and $E\,B$ (e) 
angular power spectra for $m=10^{-22}$ eV and $g_\phi=10^{-20}\,\mbox{eV}^{-1}$ 
(black solid line), the black dotted line is the standard 
case in which there is no coupling between photons and pseudoscalars ($\theta=0$). 
For the $B\,B$ power spectrum (b) we plot for comparison also 
the polarization signal induced by gravitational lensing (black dotted line), and
primordial $BB$ signal if $r=0.1$ (blue dot-dashed line).
The cosmological parameters of the flat $\Lambda CDM$ model used here are $\Omega_b \, h^2 = 0.022$,
$\Omega_c \, h^2 = 0.123$, 
$\tau=0.09$, 
$n_s=1$, $A_s=2.3\times 10^{-9}$, 
$H_0 = 100 \, h \, {\rm km \, s}^{-1} \, \mathrm{Mpc}^{-1}=72\, \mathrm{km \, s}^{-1} \, \mathrm{Mpc}^{-1}$.
} 
\label{plot::EE_3}
\end{center}
\end{figure*}

The linear polarization angle is not constant in time, 
but it oscillates with varying amplitude.
If the field represents a fraction $\Omega_\mathrm{ MAT}$ of the universe energy density, then 
the amplitude of these oscillations is:
\bea
\theta_A(\Omega_\mathrm{MAT},m,g_\phi)&=&\frac{1}{4}\sqrt{\frac{3\Omega_\mathrm{ MAT}}{\pi}}\frac{g_\phi M_\mathrm{ pl}H_0}{m} \left(\frac{1}{a_0^{3/2}}-\frac{1}{a_\mathrm{ rec}^{3/2}}\right)\nonumber\\
&\simeq& \frac{1}{4}\sqrt{\frac{3\Omega_\mathrm{ MAT}}{\pi}}\frac{g_\phi M_\mathrm{ pl}H_0}{m } z_\mathrm{ rec}^{3/2}\,.
\eea
Fixed $\Omega_\mathrm{MAT}$, 
it is possible to constraint a certain region of the $(m,g_\phi)$-plane requiring 
$\theta_A(\Omega_\mathrm{ MAT},m,g_\phi)$ to be smaller of a certain angle, 
typically of the order of few degrees (see Tab.~\ref{tab:Theta_tab}). 
The excluded region considering current limits on CMB birefringence is shown in Fig.~\ref{mg01}.

Fixed a particular value for the pseudoscalar field mass and for its coupling with photons
we can also estimate how the polarization angular power spectra are modified by a rotation of the linear
polarization plane.
We modified the source term for linear polarization in the public Boltzmann 
code CAMB \cite{Lewis:1999bs} following 
Eqs.~(\ref{DeltaE}) and (\ref{DeltaB}). 
The linear polarization rotation angle is given by Eq.~(\ref{Eq.:rot_anfle})
and the evolution of the pseudoscalar field by  Eq.~(\ref{E:phi_t_1}).
The new power spectra are compared with the standard unrotated ones 
in Fig.~\ref{plot::EE_3} 
fixed 
$m=10^{-22}$ eV and $g_\phi=10^{-20}\,\mbox{eV}^{-1}$.

In Section \ref{comparison_constantAngle} we compare the power spectra 
modified version of CAMB obtained starting by Eqs. (\ref{DeltaE},\ref{DeltaB})
which takes into account the time dependence of the pseudoscalar field 
in the integral along the line of sight
with the approximated spectra obtained following Eqs. (55-59). 

\subsection{Comments for axion cosmology}
For axions the coupling constant with photons $g_\phi$ and the energy scale
$f_a$ at which the new symmetry is broken are related \cite{Raffelt:1996wa}:
\be
\left|g_\phi\right|=\frac{\alpha_{EM}}{2 \pi f_a}\frac{3}{4}\xi\quad\mbox{with}\quad 0.1\lesssim\xi\lesssim 1\,,
\label{eq:ftog}
\ee
where the value for $\xi$ depends on the particular model considered for the axion. 
By using this relation a limit on  the coupling constant is turned into 
a limit on the energy of symmetry breaking.
 
The critical density associated with the 
misalignment production of axions strongly depends on the initial misalignment 
angle associated with the axion field $\Theta_i$
through the following relation \cite{Kolb:1990vq,Raffelt:1996wa}:
\be
\Omega_{mis}h^2\sim 0.23\times10^{\pm0.6}\left(\frac{f_a}{10^{12}\,\mbox{GeV}}\right)^{1.175}\Theta_i^2 F\left(\Theta_i\right)\,,
\ee
where $h$ encodes the actual value of the Hubble parameter ($H_0=100h\,\mbox{km}\,\mbox{s}^{-1}\,\mbox{Mpc}^{-1}$) and 
$F(\Theta_i)$ accounts for anharmonic effects if $\Theta_i\gg 1$. 
The demand $\Omega_{mis}\le\Omega_{DM}$ provides an upper bound on $f_a^{1.175}\Theta_i^2 $ (assuming $F\left(\Theta_i\right)\simeq 1$) \cite{Dine:1982ah,Abbott:1982af,Preskill:1982cy}:
\be
\label{eq:f_a_theta}
f_a \Theta_i^{1.7}\le2\times10^{11 \div 12}\,\mbox{GeV}\,.
\ee

This condition becomes also an upper bound for $f_a$ under the assumption 
that inflation occurred before the breaking of PQ-symmetry ($f_a\le f_{INF}$) 
\cite{Kolb:1990vq}: in this scenario 
different regions have different values for $\Theta_i$, so averaging over 
all observable universe 
the value of $\Theta_i$ in equation can replaced by its $rms$ value 
($\pi/\sqrt{3}$) 
and the limit $f_a\le10^{11 \div 12}\,\mbox{GeV}$ is obtained. 
As can be seen from Fig. 2, CAST disfavors values of $g_\phi \sim 
10^{-11 \div -12}\,\mbox{GeV}^{-1}$ with a mass up to $0.02 \mbox{eV}$. 
Note however that our calculation cannot be applied directly to this 
case since we assume $\phi$ homogeneous in our universe, whereas it is not 
if the PQ symmetry breaking occurs after inflation: although taking 
into account space inhomogeneities were a second order effect in cosmological 
perturbation theory, cosmological birefringence might be larger than the one 
computed in this paper.

Our calculations apply without modifications to the case in which inflation 
occurs after PQ-symmetry breaking: the initial misalignment angle $\Theta_i$
is homogeneous throughout our universe and can be much smaller than 
$\pi/\sqrt{3}$. Such possibility allow the scale of PQ-symmetry breaking 
$f_a$ to be much higher than $10^{11 \div 12}\,\mbox{GeV}$ and 
is motivated by anthropic 
considerations \cite{Pi:1984pv,Linde:1987bx,Tegmark:2005dy,Hertzberg:2008wr}. 
These smaller values of $g_\phi$ can be constrained by present data in CMB 
polarization in a much better way than CAST, in particular for small masses.

\section{Exponential potential}
We consider in this section a pseudoscalar field with an exponential potential:
\begin{equation}
V\left(\phi\right)=V_0 \exp\left(-\lambda \kappa \phi\right)\,,
\end{equation}
with $\kappa^2\equiv8 \pi G\,.$ 
Theoretical motivations to this kind of potential are certainly weaker than the ones 
for the potential presented in Eq. (\ref{alps_pot}).
However it is interesting to show how the kinematics of the pseudoscalar field 
is important for the resulting spectra of CMB anisotropies in polarization. 
Whereas the time derivative of the pseudoscalar field in the previous case contains 
oscillations about a vanishing value (see Eq. (\ref{phip_1})),
we  study here a case where the behaviour is monotonous. 

It is known \cite{Copeland:1997et} that exponential potential 
with $\lambda^2>3(1+w_{\rm F})$ leads to a component which tracks the dominant 
background fluid with equation of state 
$p_{\phi}=w_{\phi} \rho_{\phi}$. 
In order to satisfy the nucleosynthesis bound we choose $\lambda=4.5$. 
During the matter dominated era the scalar field behaves as:
\begin{eqnarray}
\rho_\phi&=&\frac{\dot{\phi}^2}{2}+V_0 \exp\left(-\lambda \kappa \phi\right)=f\, \rho_{\rm MAT}\equiv
f \frac{\rho_{\rm MAT,0}}{a^3},\\
P_\phi&=&\frac{\dot{\phi}^2}{2}-V_0 \exp\left(-\lambda \kappa \phi\right),
\end{eqnarray}
where $\rho_{\rm MAT}=\rho_{\rm DM}+\rho_{\rm baryons}+\rho_\phi$.

\begin{figure}
\begin{tabular}{c}
\includegraphics[width=8.6cm]{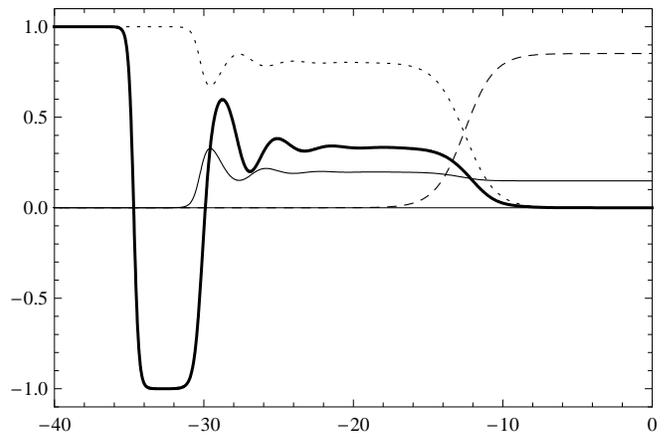}
\end{tabular}
\caption{For $\lambda=4.5$ Dashed line: $\Omega_{\rm DM}+\Omega_{\rm baryons}$, 
dotted line: $\Omega_{\rm RAD}$, 
thin continuous line: $\Omega_{\rm \phi}$, thick continuous line: $w_{\rm \phi}$, 
in terms of the natural logarithm of the scale factor 
(from $\log a\simeq-40$ to nowadays $\log a_0=0$). 
Here $\Omega_{\rm DM,0}+\Omega_{\rm baryons,0}=0.852$ and $\Omega_{\rm \phi,0}=0.148$.}
\label{fig_EXP1}
\end{figure}

For $\lambda=4.5$ the contribution of the pseudoscalar field to universe energy density is shown in 
Fig.~\ref{fig_EXP1}. The value of $\Omega_\phi$ changes with time, 
but it is almost constant ($\Omega_\phi\simeq \Omega_{\phi\,,0}=0.148$) 
from recombination ($\log a_{\rm rec}\simeq-7$) to nowadays.

The derivative of the pseudoscalar field respect to conformal time is proportional to $a^{-1/2}$ and 
the evolution of the scale factor in the matter dominated phase is $a (\eta) = (\eta/\eta_0)^2$ 
so:
\begin{equation}
\phi^{\prime} =\sqrt{f\, \rho_{\rm MAT, 0}}\frac{\eta_0}{\eta} \,.
\end{equation}
Substituting this relation in Eq.~(\ref{eq:A}) we obtain the following expression for the evolution 
of the electromagnetic potential:
\begin{equation}
\tilde{A}_{\pm}^{\prime\prime}+\left( k^2 \pm g_\phi\sqrt{f\, \rho_{\rm MAT, 0}} \frac{\eta_0}{\eta} k \right)
 \tilde{A}_{\pm}=0\,.
\end{equation}  
This is a particular differential equation, called Coulomb wave equation; 
defining $q_{\pm}\equiv \mp g_\phi\sqrt{f\,\rho_{\rm MAT ,0}}\eta_0/2 =\mp q $ and $x\equiv k\eta$ it becomes:
\begin{equation}
\frac{d^2\tilde{A}_{\pm}}{dx^2}+\left(1-\frac{2q_{\pm}}{x} \right) \tilde{A}_{\pm}=0\,.
\label{eq:Coul}
\end{equation} 
The solution of this particular equation can be written in terms of regular ($F_0(q,x)$) and 
irregular ($G_0(q,x)$) Coulomb wave functions \cite{Abramowitz,Anber:2006xt}: 
\begin{eqnarray*}
\tilde{A}_+&=&f_+ F_0(q_+,x)+g_+ G_0(q_+,x)\\
&=&f_+ F_0(-q,x)+g_+ G_0(-q,x)\,,\\
\tilde{A}_-&=&f_- F_0(q_-,x)+g_- G_0(q_-,x)\\
&=&f_- F_0(q,x)+g_- G_0(q,x)\,,
\end{eqnarray*}
where $f_+,f_-,g_+,g_-\in \mathbb{C}$; in a compact notation:
\begin{eqnarray}
\label{eq:Coul3}
\tilde{A}_\pm (q,x)&=&f_\pm F_0(\mp q,x)+g_\pm G_0(\mp q,x)\,.
\end{eqnarray}
The Stokes parameters contain the derivative respect to conformal time $\eta$, so we evaluate:
 \begin{eqnarray}
\tilde{A}_\pm^{\prime}(q,x)&=&k\left[ f_\pm \frac{\partial F_0(\mp q,x)}{\partial x}+g_\pm \frac{\partial G_0(\mp q,x)}{\partial\, x}\right]\,.
\end{eqnarray}
The solution given in Eq.~(\ref{eq:Coul3}) verifies the {\it Wronskian condition} 
($\tilde{A}_\pm^{}\tilde{A}_\pm^{\prime *}-\tilde{A}_\pm^{\prime}\tilde{A}_\pm^{*}=i$) 
if the following relation holds:
\begin{equation}
\label{Wrons}
f_{\pm}^{*}g_{\pm}-f_{\pm}g_{\pm}^{*}=
\frac{i}{k}\quad\Longrightarrow\quad\Im\left(f_\pm^*g_\pm\right)=
\frac{1}{2k}\,.
\end{equation}

In the general case, when the coupling does not vanishes ($g_{\phi}\neq0$),
 we expand the solution~(\ref{eq:Coul3}) for large value of $x$ 
neglecting terms proportional to $\mathcal{O}(x^{-2})$ (see Appendix): 
\begin{widetext}
\begin{eqnarray}
\label{eq:A2}
\tilde{A}_\pm(q,x)&\simeq& f_\pm \left[\frac{q^2}{2 x}\cos \left(x\pm\alpha\left(q,x\right)\right)+\left(1\mp \frac{q}{2 x}\right) \sin \left(x\pm\alpha\left(q,x\right)\right) \right] \nonumber \\
 & &  +g_\pm \left[ \left( 1\mp \frac{q}{2 x} \right) \cos\left(x\pm\alpha\left(q,x\right)\right) -\frac{q^2}{2 x}\sin \left(x\pm\alpha\left(q,x\right)\right) \right]\,,
\end{eqnarray}
where $\alpha\left(q,x\right)\equiv q \ln 2 x -\arg \Gamma(1+ iq)$. 
The derivative respect to conformal time is:
\begin{eqnarray}
\label{eq:dA}
\tilde{A}_\pm^{\prime}(q,x)&\simeq& k\left\{ f_\pm \left[\left(1\pm \frac{q}{2 x}\right)\cos \left(x\pm\alpha\left(q,x\right)\right)-\frac{q^2}{2 x} \sin \left(x\pm\alpha\left(q,x\right)\right) \right] \right. \nonumber \\
 & & \left.+g_\pm \left[-\frac{q^2}{2 x} \cos\left(x\pm\alpha\left(q,x\right)\right) - \left(1\pm \frac{q}{2 x}\right)\sin \left(x\pm\alpha\left(q,x\right)\right) \right]\right\}\nonumber\\
& & = \frac{k}{2}\left[e^{i \left(x\pm\alpha\left(q,x\right)\right) }
\left(1\pm\frac{q}{2x}+i\frac{q^2}{2x}\right)\left(f_\pm+i g_\pm\right)\right.\nonumber\\
& & \left. e^{-i \left(x\pm\alpha\left(q,x\right)\right) }
\left(1\pm\frac{q}{2x}-i\frac{q^2}{2x}\right)\left(f_\pm-i g_\pm\right)  \right]\,.
\label{eq:A3}
\end{eqnarray}
\end{widetext}
In general both forward moving waves ($\tilde{A}_\pm\propto e^{-ik\eta}$) 
and backward moving waves ($\tilde{A}_\pm\propto e^{ik\eta}$) must be taken into account 
for propagation of light in a medium. 
Chosen a particular value for the constants $f_\pm$ and $g_\pm$ 
that verifies the Wronskian relation~(\ref{Wrons}) the evolution of polarization is fixed. 

If we assume, according with \cite{Jain:2002vx,Mirizzi:2006zy}, 
that the photon pseudoscalar conversion is a small effect due to low 
energy of CMB photons, the production of backward moving waves 
can be neglected (see \cite{Das:2004qk} for the use of this approximation). 
The Eq.~(\ref{eq:A3}) setting $f_\pm=-i g_\pm$ becomes: 
\begin{equation}
\tilde{A}^{\prime}_\pm\left(q,x\right)\simeq -i k g_\pm \left(1\pm\frac{q}{2 x}-i\frac{q^2}{2 x}\right)
e^{-i\left(x\pm\alpha(q,x)\right)} \,,
\end{equation}
and in terms of the value at recombination time:
\begin{eqnarray}
& & \tilde{A}^{\prime}_\pm\left(q,x\right)\simeq \tilde{A}^{\prime}_\pm\left(q,x_{\rm rec}\right)
\left[1\pm\frac{q}{2}\left(\frac{1}{x}-\frac{1}{x_{\rm rec}}\right)\right.\nonumber\\
& & \left. -i\frac{q^2}{2}\left(\frac{1}{x}-\frac{1}{x_{\rm rec}}\right)\right] \exp\left\{-i\left[x-x_{\rm rec}\pm\Delta\alpha\right]\right\}\,,
\end{eqnarray}
where we have introduced  
\bea
\Delta\alpha&\equiv& \alpha(q,x)-\alpha(q,x_{\rm rec})
=q\ln\left(\eta/\eta_{\rm rec}\right)\nonumber\\
&=& \frac{q}{2}\ln\left(a/a_{\rm rec}\right) \,.
\eea
We observe that also in this exact case the plane of linear polarization is rotated of 
an angle $\Delta\alpha$ independent on $k$ whose dependence on 
the difference between the present value of $\phi$ and the corresponding 
one at recombination is the same of the adiabatic approximation and of 
Eq. (\ref{assumption}). 


\begin{figure*}
\begin{center}
\begin{tabular}{c}
\includegraphics[width=8.6cm]{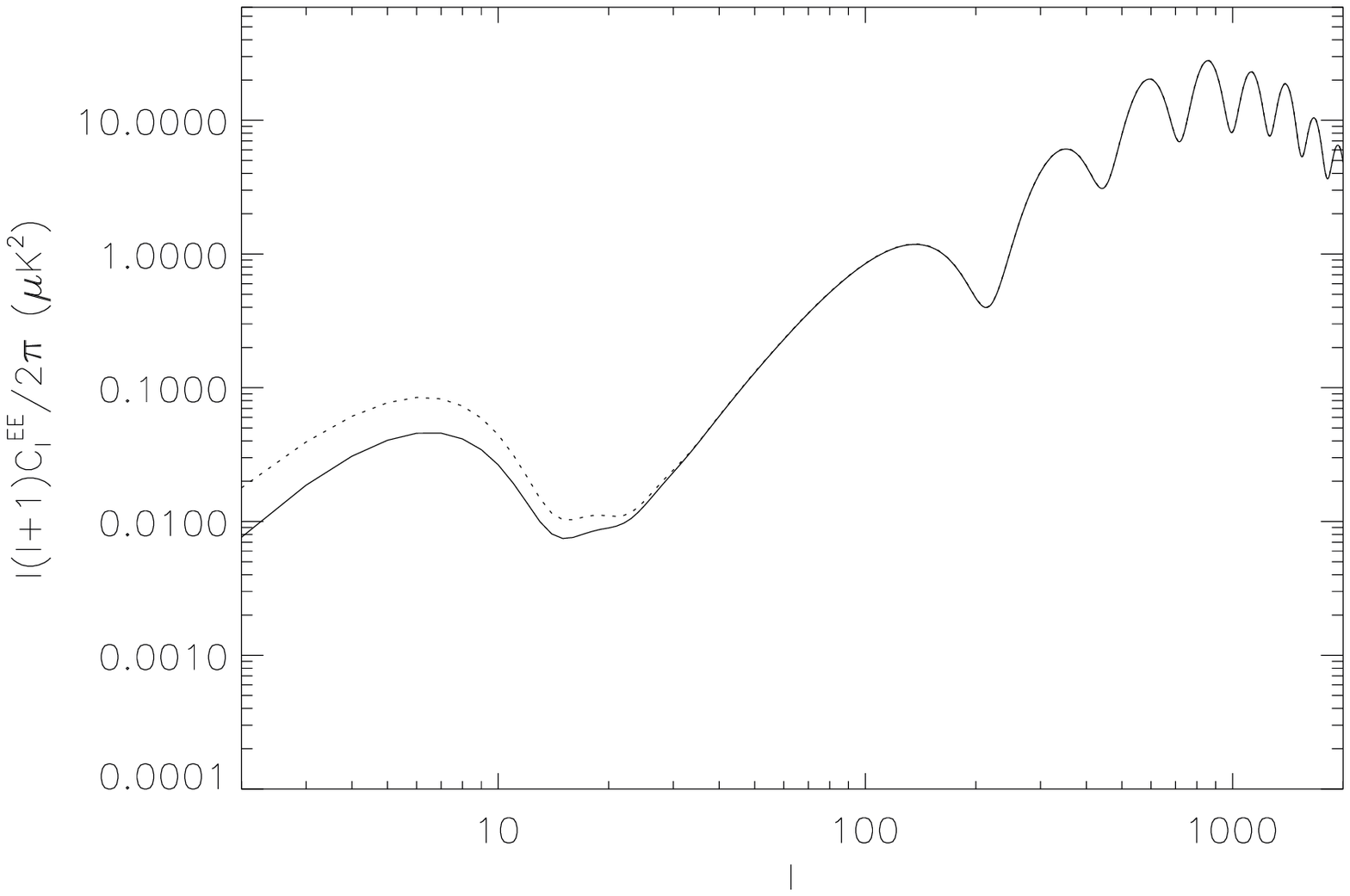}
\includegraphics[width=8.6cm]{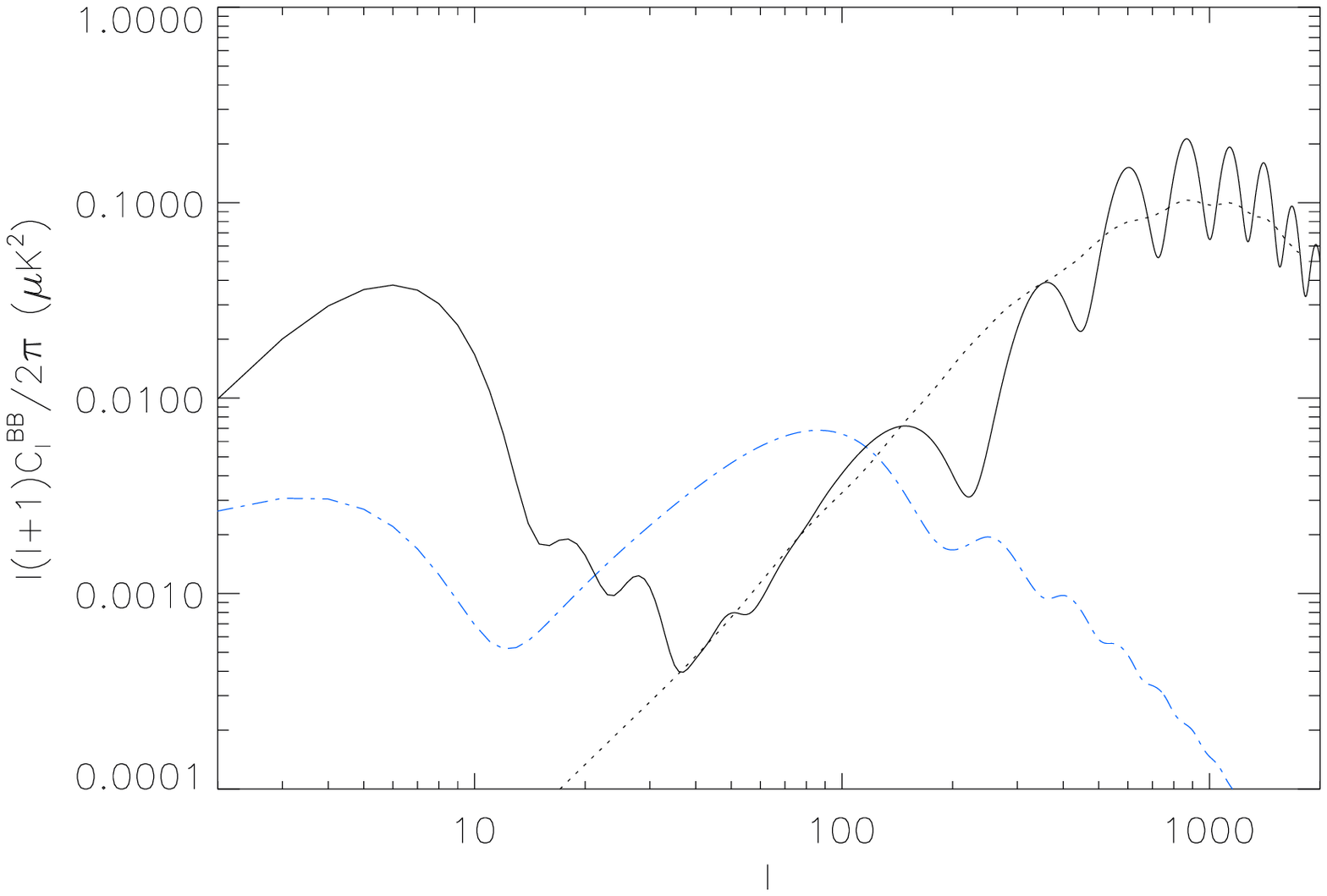}\\
\includegraphics[width=8.6cm]{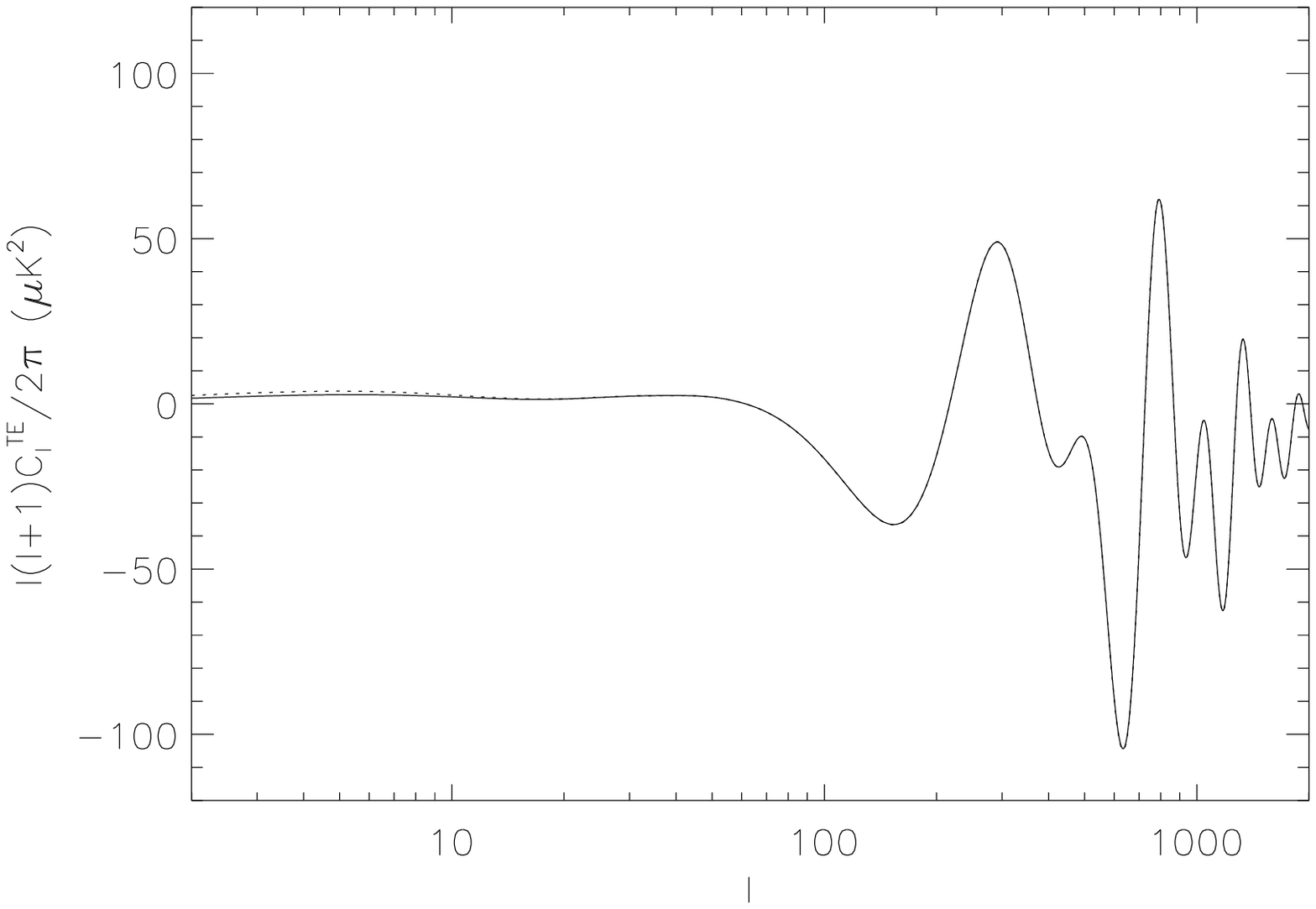}
\includegraphics[width=8.6cm]{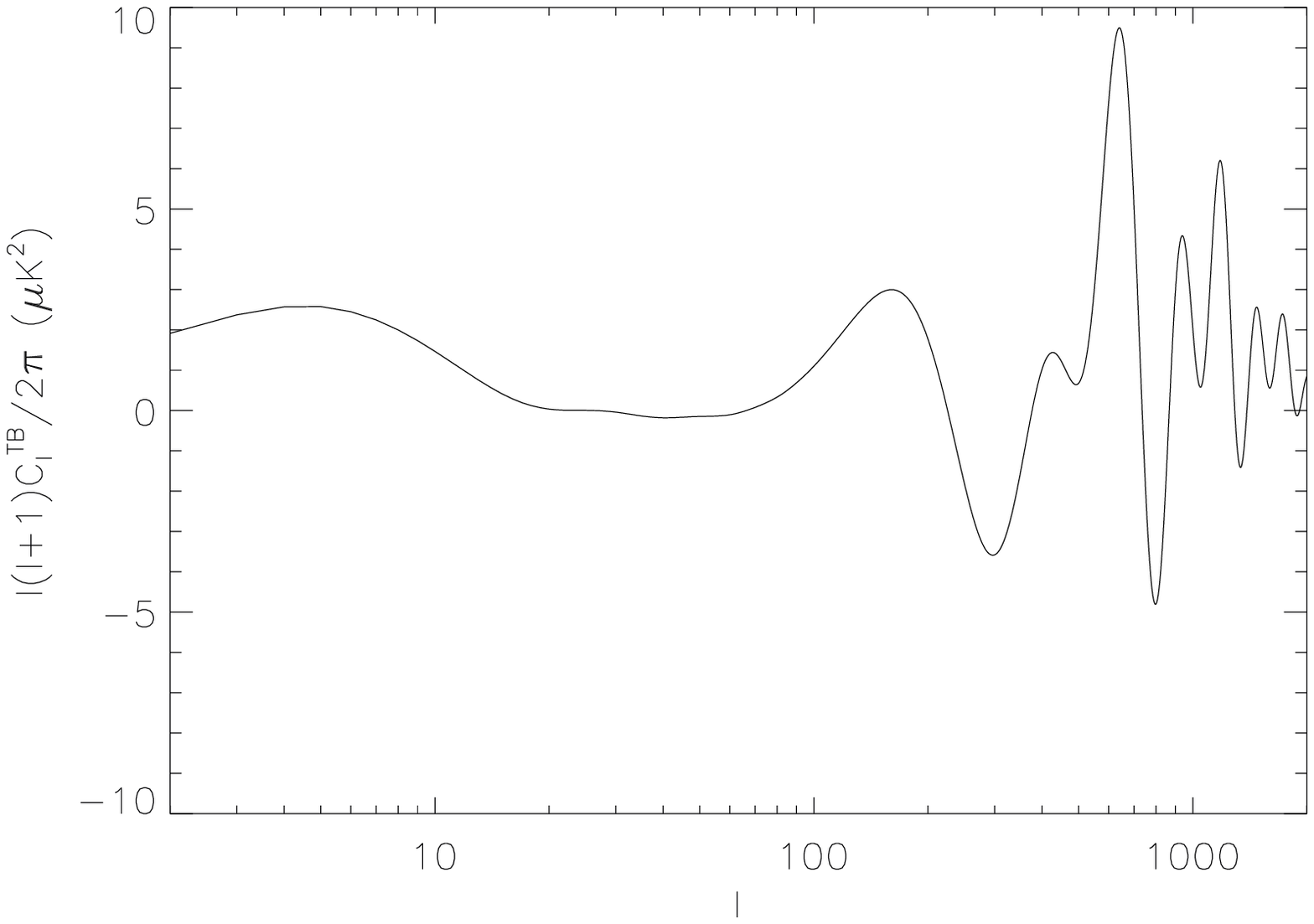}\\
\includegraphics[width=8.6cm]{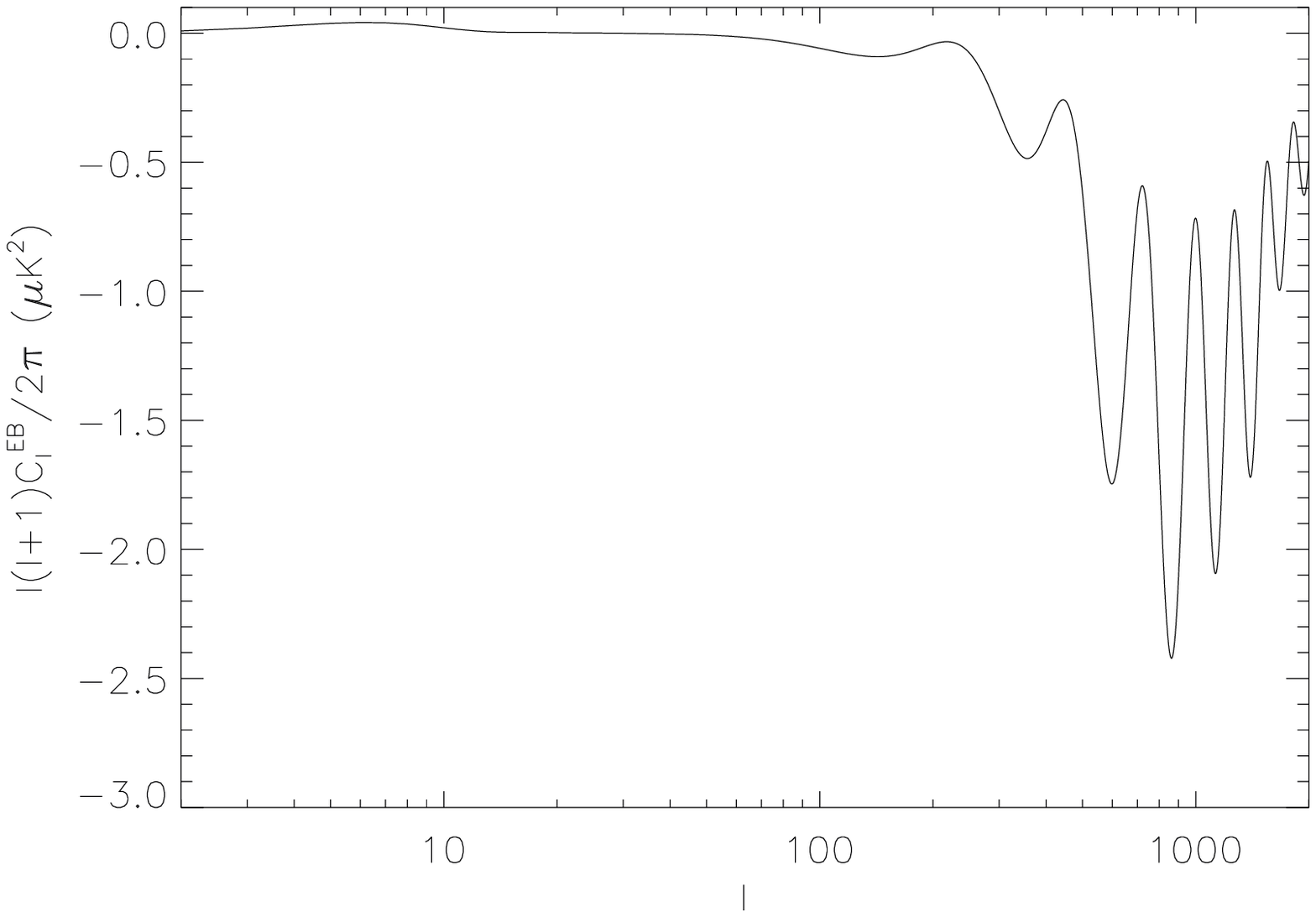}
\end{tabular}
\caption{$E\,E$ (a), $B\,B$ (b), $T\,E$ (c), $T\,B$ (d) and $E\,B$ (e)
angular power spectra for $g_\phi=10^{-28}\,\mbox{eV}^{-1}$ 
(black solid line); the black dotted line is the standard 
case in which there is no coupling ($\theta=0$). 
For the $B\,B$ power spectrum (b) we plot for comparison also 
the polarization signal induced by gravitational lensing (black dotted line), and
primordial $BB$ signal if $r=0.1$ (blue dot-dashed line).
The cosmological parameters of the flat $CDM$ model used here are $\Omega_b= 0.0462$,
$\Omega_c = 0.9538$ ($\Omega_\phi\simeq 0.148$), 
$\tau= 0.09$, 
$n_s=1$, $A_s=2.3\times 10^{-9}$, 
$H_0 = 72\, \mathrm{km \, s}^{-1} \, \mathrm{Mpc}^{-1}$.
}
\label{Bolt_mono_8}
\end{center}
\end{figure*}

Current measures and constraints on the 
polarization pattern of CMB anisotropies produce an upper limit on the 
linear polarization rotation angle of the order of few degrees 
(see Tab.~\ref{tab:Theta_tab}).
We now use these constraints and our analytic expression:
\bea
 \left|\theta\right|&=& \frac{|q|}{2} \ln (1 + z_\mathrm{ rec} )\nonumber\\
 &\simeq&\frac{1}{4}\sqrt{\frac{3}{2\pi}\Omega_{\phi,0}} \left|g_\phi\right| M_\mathrm{ pl } \ln (1 + z_\mathrm{ rec} )
 \,,
\eea
to obtain an upper bound for $q$, which can be turned into a upper bound on $g_\phi$;
if $\left|\theta\right|\lesssim 6$ deg, then:
\begin{equation}
\left|g_\phi\right| \lesssim 10^{-30} \,\mbox{eV}^{-1}\,,
\end{equation}
where we have assumed: $\Omega_{\phi,0}\simeq 0.148$ and $z_\mathrm{ rec}\simeq 1100$.

The angle of linear polarization $\theta(\eta)$ appearing 
in Eqs.~(\ref{DeltaE}) and (\ref{DeltaB})
can be replaced with: 
\be
\left| \theta(\eta)\right|\simeq\frac{1}{2}\sqrt{\frac{3}{2\pi}\Omega_{\phi,0}} \left|g_\phi\right| M_\mathrm{ pl } \ln \left(\frac{\eta}{\eta_\mathrm{rec}} \right)\,,
\ee
and the polarization power spectra are evaluated using the expression given in Section \ref{sect_Bolz_a},
see angular power spectra of Fig.~\ref{Bolt_mono_8}.

In Section \ref{comparison_constantAngle} we compare the power spectra
modified version of CAMB obtained starting by Eqs. (\ref{DeltaE},\ref{DeltaB})
which takes into account the time dependence of the pseudoscalar field
in the integral along the line of sight
with the approximated spectra obtained following Eqs. (55-60).

\section{Comparison with constant rotation angle approximation}
\label{comparison_constantAngle}

\begin{figure*}
\begin{center}
\begin{tabular}{c}
\includegraphics[width=8.6cm]{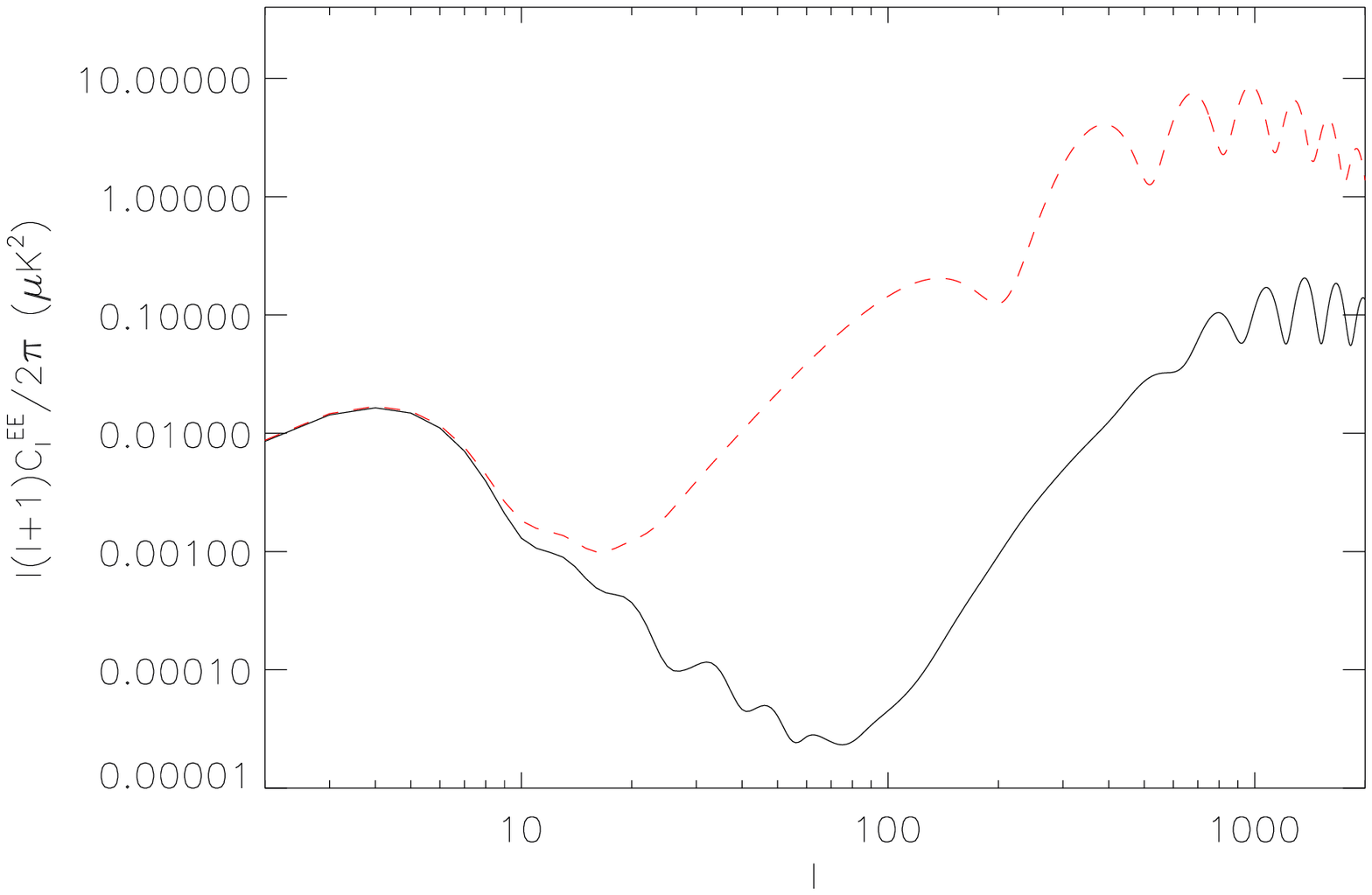}
\includegraphics[width=8.6cm]{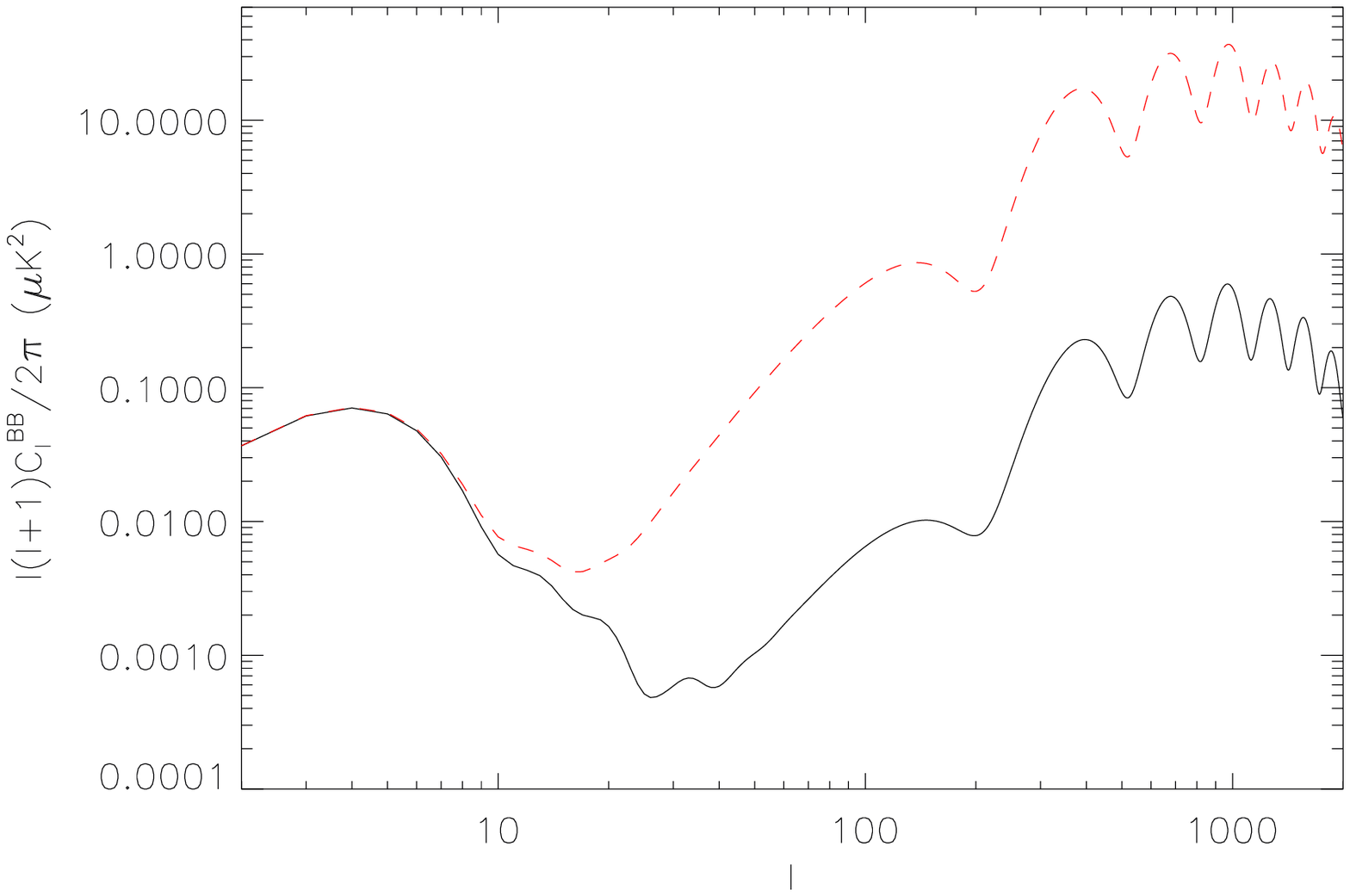}\\
\includegraphics[width=8.6cm]{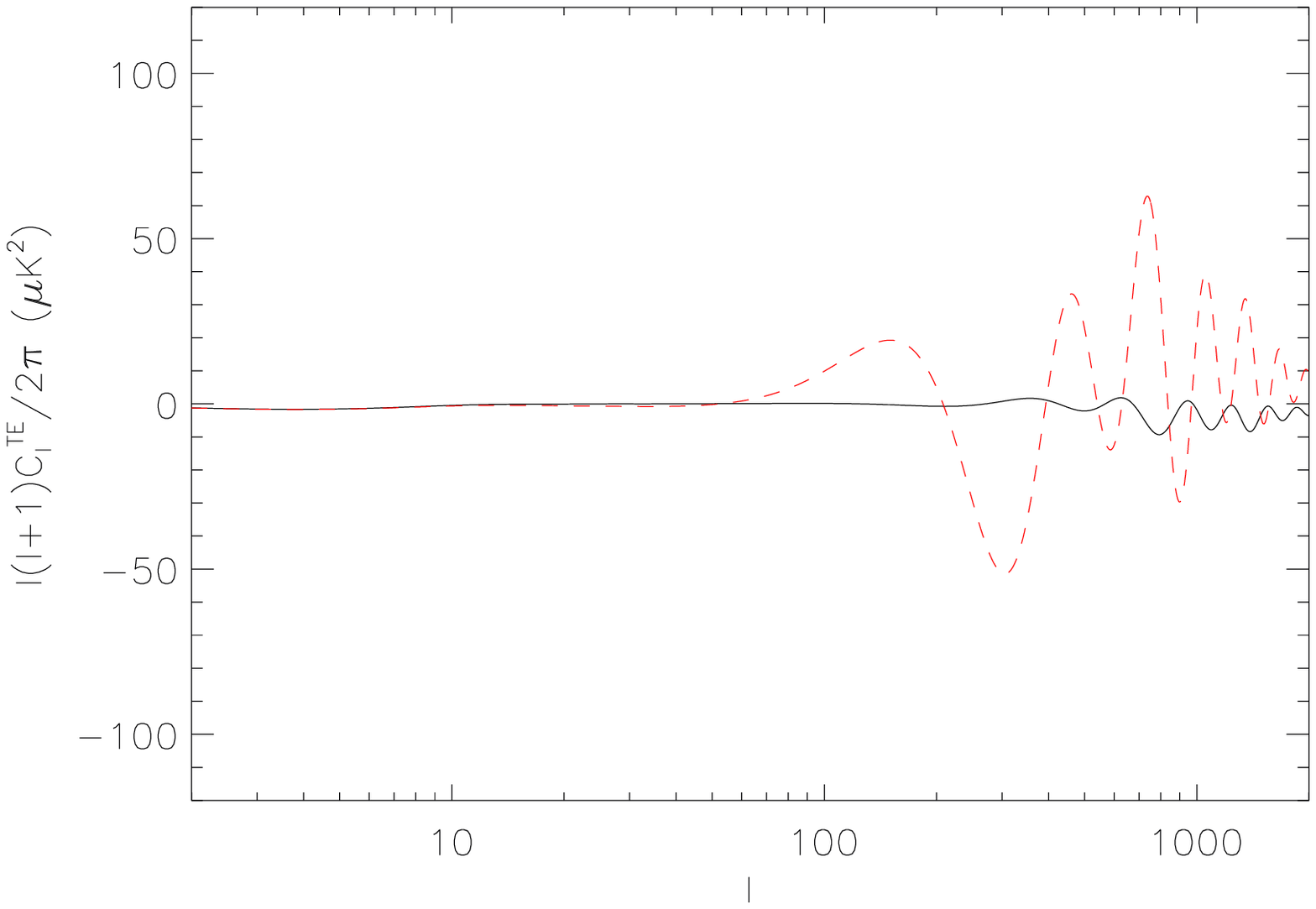}
\includegraphics[width=8.6cm]{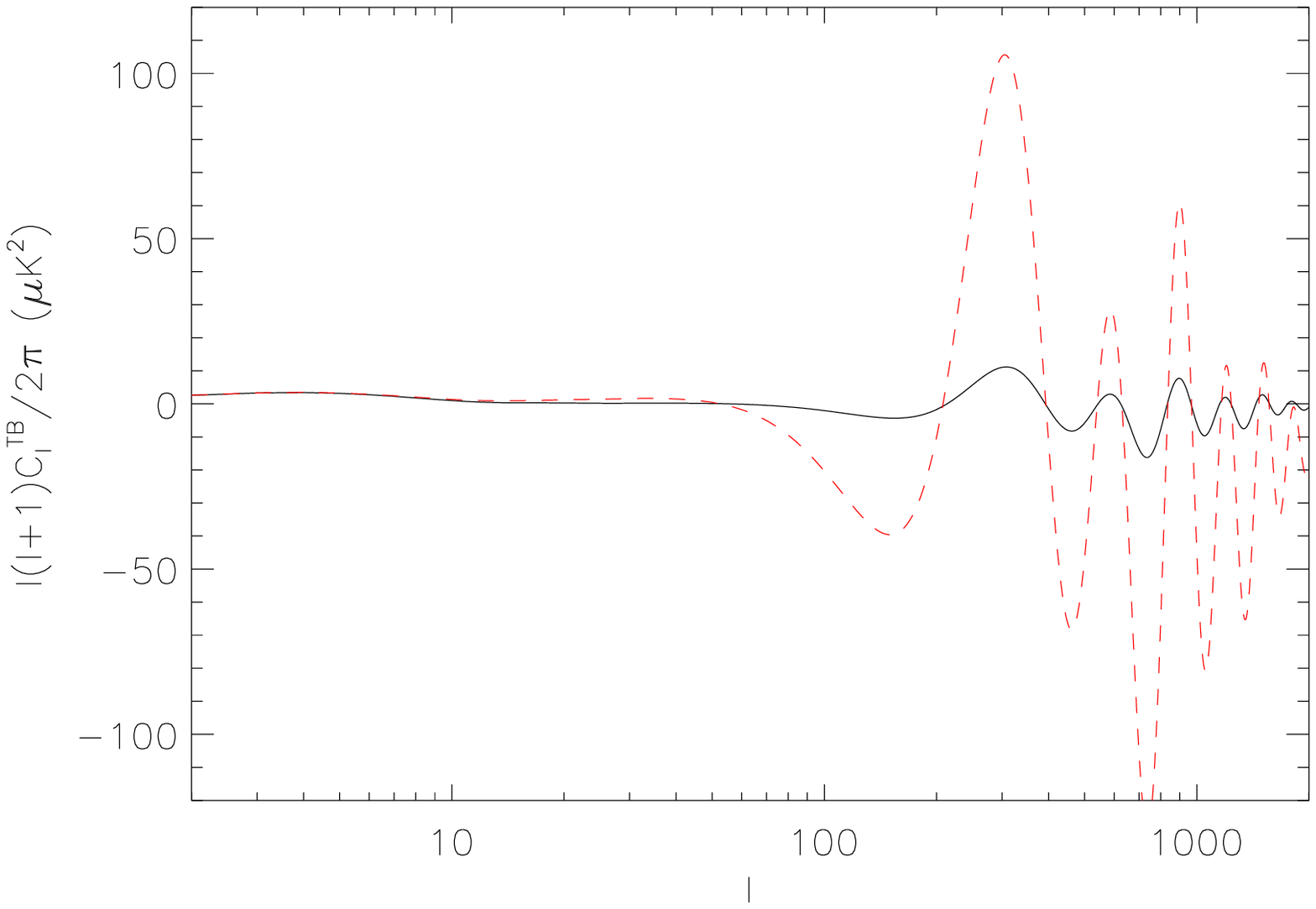}\\
\includegraphics[width=8.6cm]{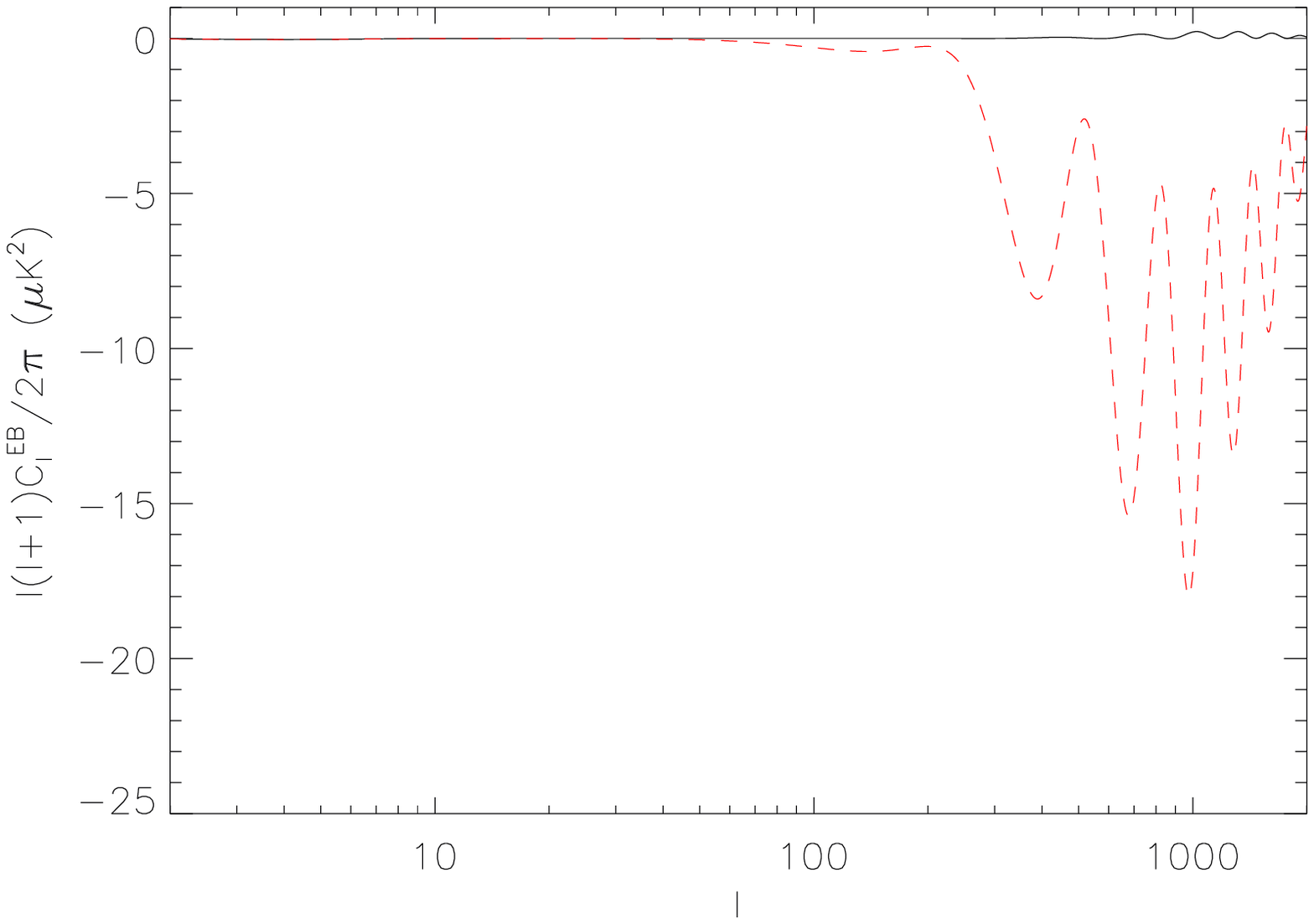}
\end{tabular}
\caption{$E\,E$ (a), $B\,B$ (b), $T\,E$ (c), $T\,B$ (d) and $E\,B$ (e) 
angular power spectra for $m=10^{-22}$ eV and $g_\phi=10^{-20}\,\mbox{eV}^{-1}$ 
(black solid line) and approximating the rotation angle with the constant 
value $\theta_\mathrm{rec}$ (red dashed line).
The cosmological parameters of the flat $\Lambda CDM$ model used here are $\Omega_b \, h^2 = 0.022$,
$\Omega_c \, h^2 = 0.123$, 
$\tau=0.09$, 
$n_s=1$, $A_s=2.3\times 10^{-9}$, 
$H_0 = 100 \, h \, {\rm km \, s}^{-1} \, \mathrm{Mpc}^{-1}=72\, \mathrm{km \, s}^{-1} \, \mathrm{Mpc}^{-1}$.
} 
\label{plot::EE2_4}
\end{center}
\end{figure*}

\begin{figure*}
\begin{center}
\begin{tabular}{c}
\includegraphics[width=8.6cm]{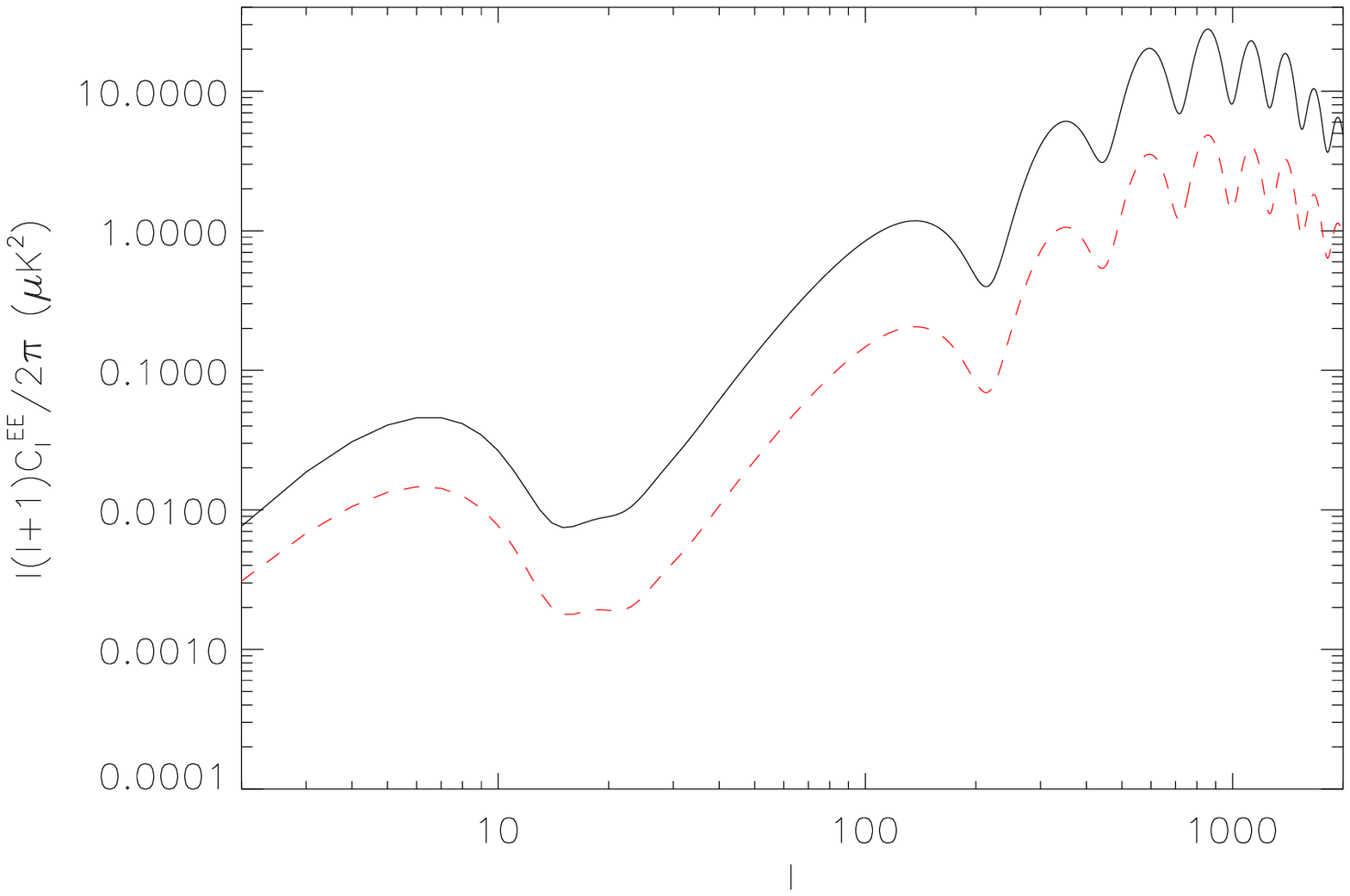}
\includegraphics[width=8.6cm]{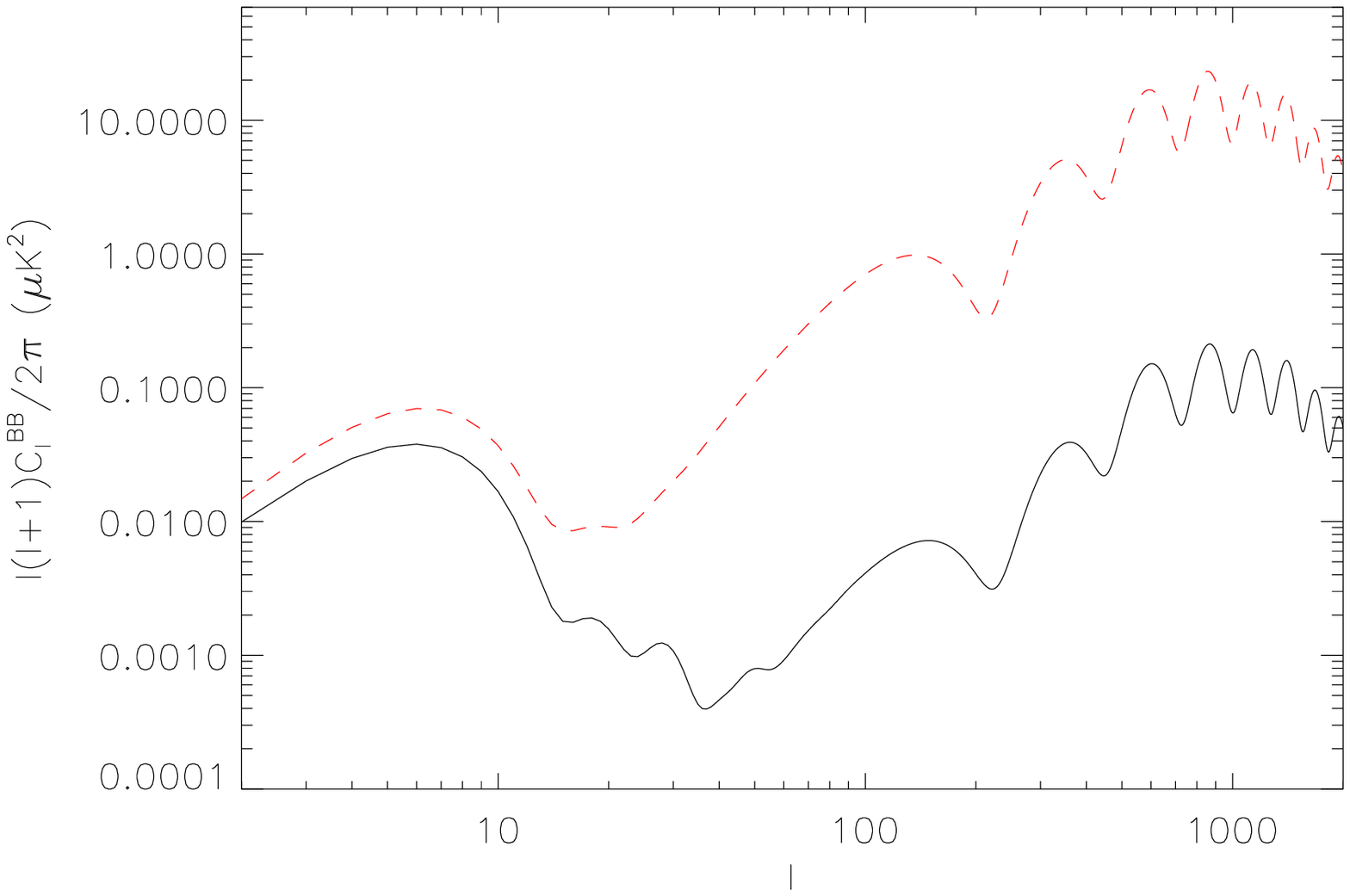}\\
\includegraphics[width=8.6cm]{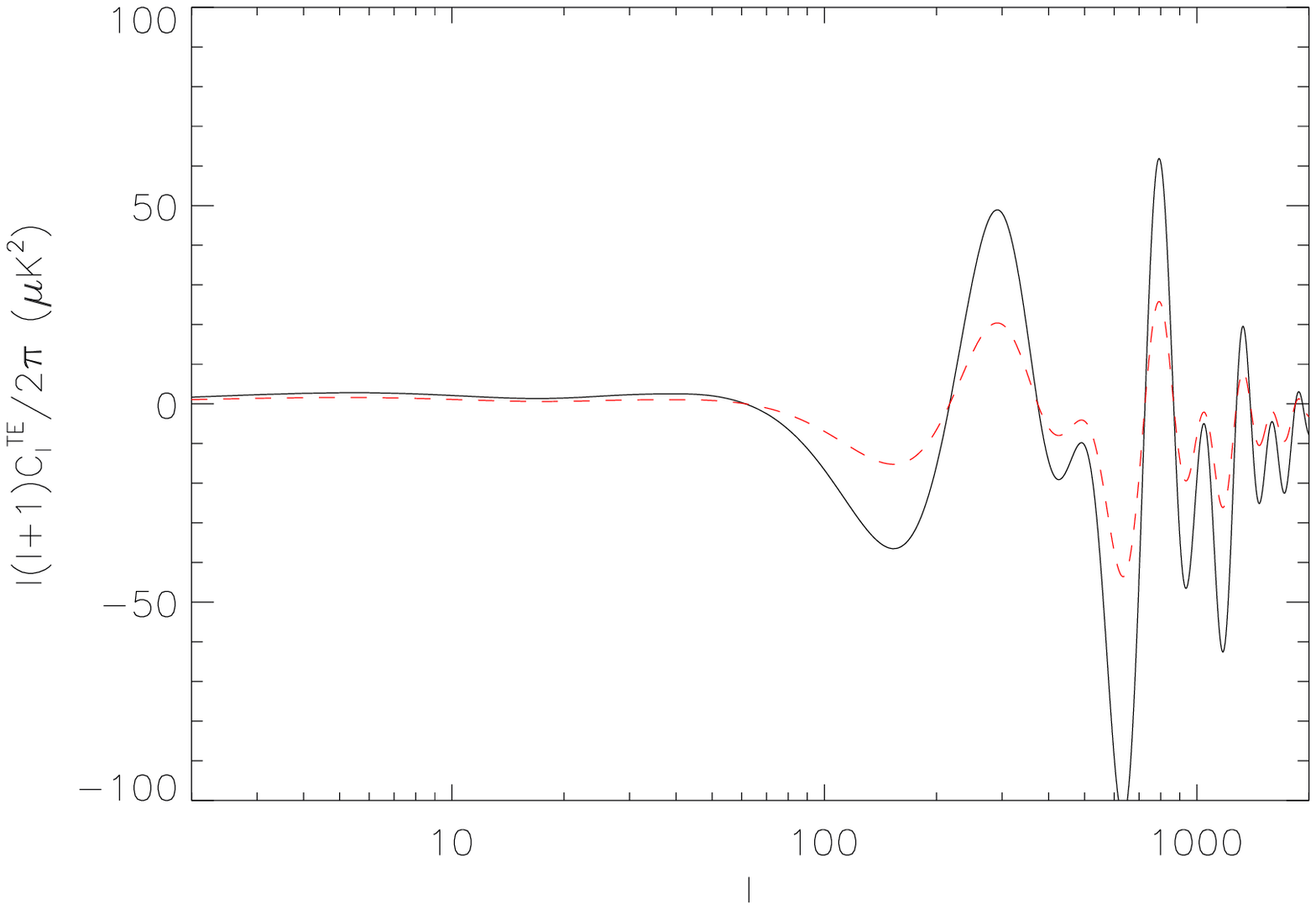}
\includegraphics[width=8.6cm]{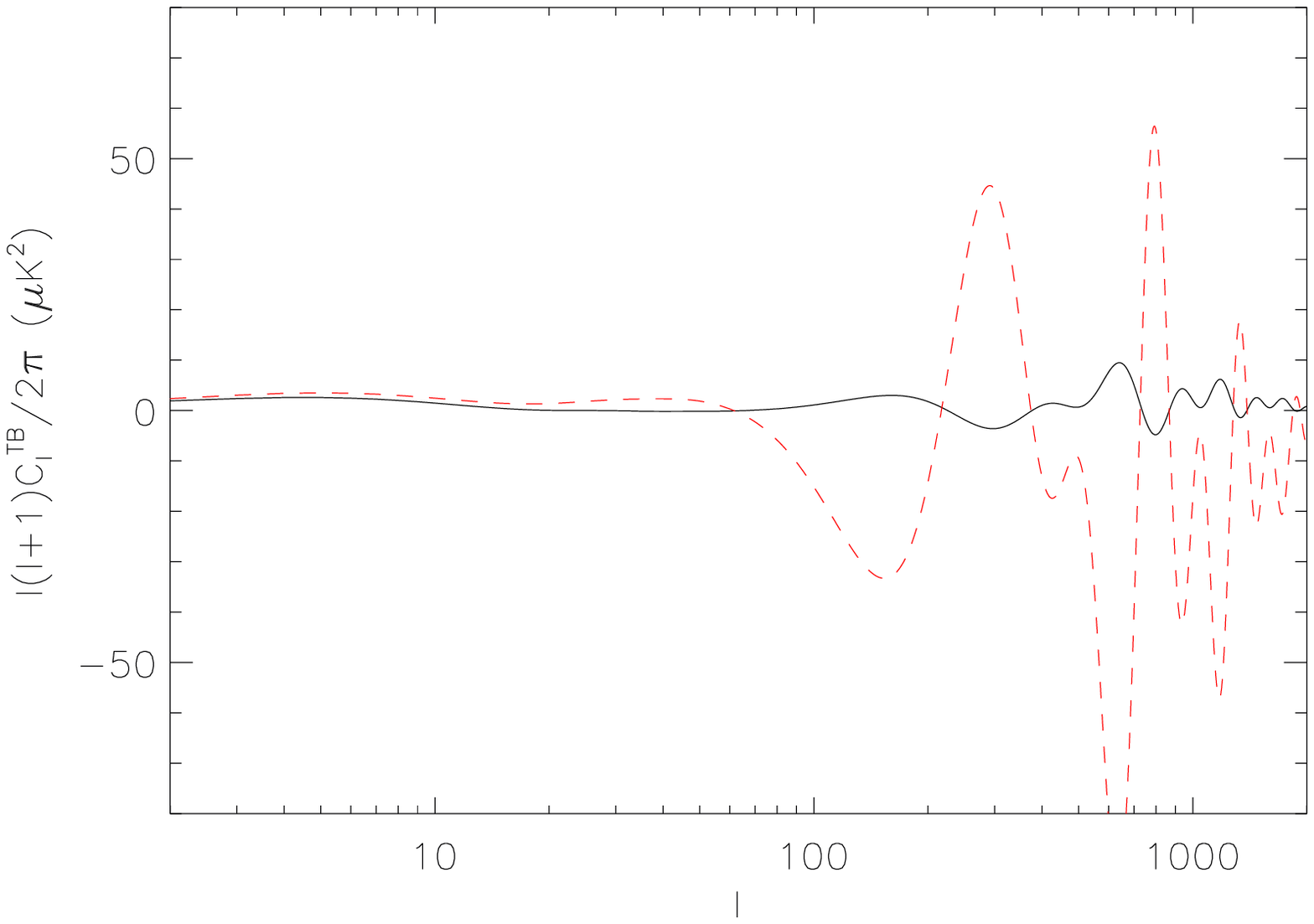}\\
\includegraphics[width=8.6cm]{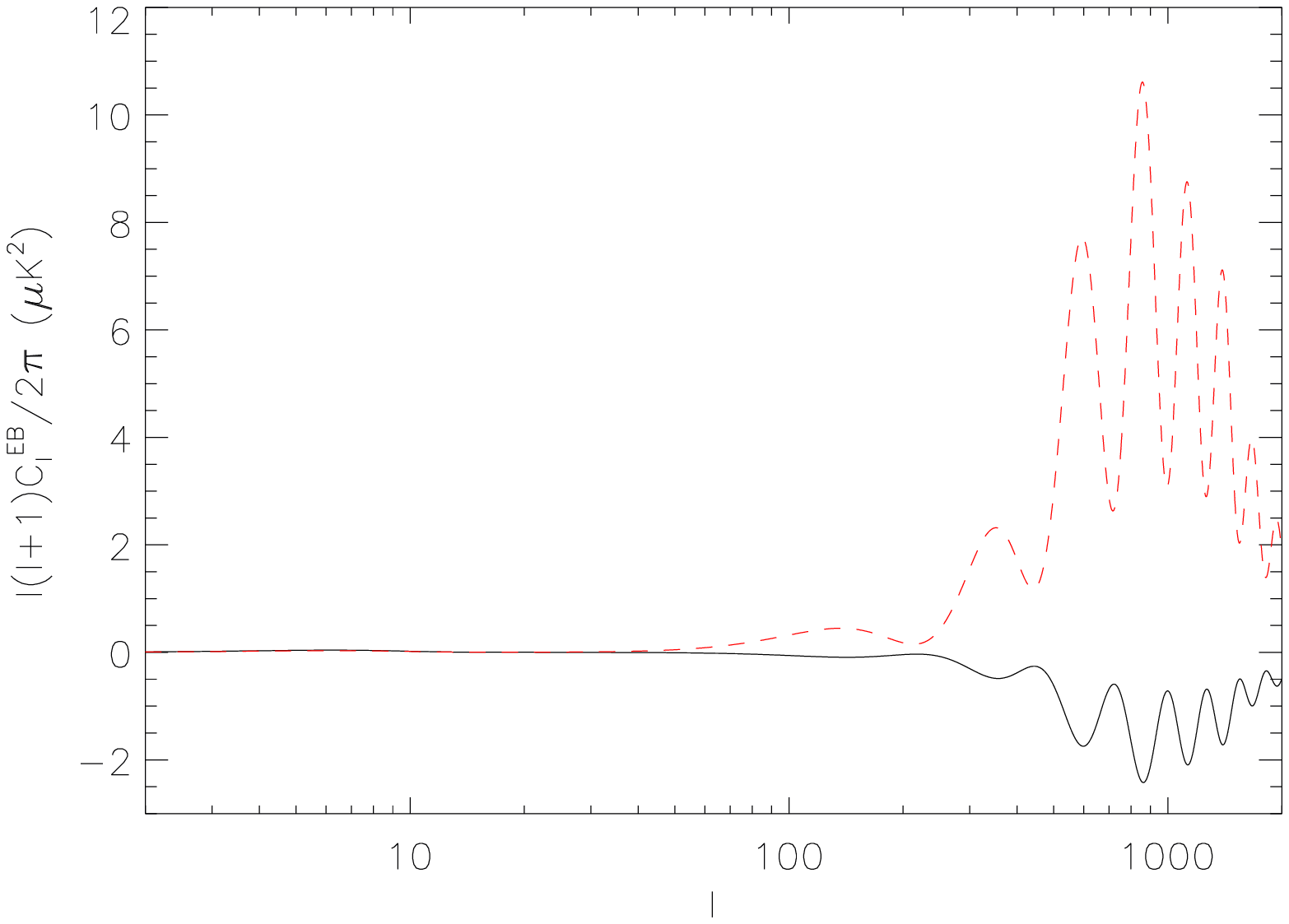}
\end{tabular}
\caption{$E\,E$ (a), $B\,B$ (b), $T\,E$ (c), $T\,B$ (d) and $E\,B$ (e) 
angular power spectra for $g_\phi=10^{-28}\,\mbox{eV}^{-1}$ 
(black solid line) and approximating the rotation angle with the constant 
value $\theta_\mathrm{rec}$ (red dashed line); the black dotted line is the standard 
case in which there is no coupling. 
The cosmological parameters of the flat $CDM$ model used here are $\Omega_b= 0.0462$,
$\Omega_c = 0.9538$ ($\Omega_\phi\simeq 0.148$), 
$\tau= 0.09$, 
$n_s=1$, $A_s=2.3\times 10^{-9}$, 
$H_0 = 72\, \mathrm{km \, s}^{-1} \, \mathrm{Mpc}^{-1}$.
}
\label{Bolt_mono_2}
\end{center}
\end{figure*}

In this section we compare the angular power spectra obtained modifying the public code CAMB \cite{Lewis:1999bs}
considering the correct dynamic of the pseudoscalar field ($\theta=\theta(\eta)$) 
as described in Section~\ref{sect_Bolz_a}, with 
the ones obtained in the constant rotation angle approximation ($\theta=\mbox{const}$) 
for the two different potential considered in the previous sections: see Figs.~\ref{plot::EE2_4} 
and \ref{Bolt_mono_2}.

In Section~\ref{sect_Bolz_a} we have already shown how the power spectra in the constant rotation 
angle approximation [Eqs.~(\ref{C_ll_EE_constant})-(\ref{C_ll_TB_constant})] can be obtained from the general expressions [Eqs.~(\ref{C_ll_EE_dyn})-(\ref{C_ll_TB_dyn})].   

Power spectrum  modifications obtained starting directly form the Boltzmann equations and 
taking into account the temporal evolution of the pseudoscalar field are usually
smaller than effects predicted considering a constant rotation angle equal to the total rotation angle 
from last scattering to nowadays.
If the cosmological pseudoscalar field evolves quickly, then the constant rotation angle approximation 
clearly leads to an overestimate of the effects. 

It is important to stress that the constant rotation angle approximation is 
just an operative approximation.
The additional term in the Boltzmann equations which rotates  
the linear polarization plane is (see Eq.~\ref{qeu}):
\be
\mp i 2\theta^{\prime}(\eta) \Delta_{Q\pm iU}(k,\eta)\,,
\ee
which clearly vanishes for $\theta=\mbox{const}$.

\section{Conclusions} 

We have studied the impact of a pseudoscalar field acting as dark matter 
on CMBP. We have shown that such pseudoscalar interaction with photons 
rotates the plane of linear polarization and generates circular polarization.
In absence of measures for the $V$ mode of CMBP, the existing upper limits on an isotropic 
$T \, B$ and $E \, B$ correlations can constrain the coupling constants 
of photons with the pseudoscalar field. 

We have examined two representative examples for the dynamics of a 
pseudo-Goldstone field 
behaving as dark matter: the oscillating and the monotonic decreasing behavior. 
In the monotonic decreasing behavior, by neglecting backward moving waves, 
we have shown how present CMB observations can constrain the coupling 
constant $g_\phi$ to  small values as  $\mathcal{O}\left(10^{-30}\right)$ eV.
For the more physically motivated axion case which leads to an oscillating 
behaviour, we have shown how constraints from CMB cosmological birefringence  
can become important for small masses for the axion.

We have also shown how the use of integral solution of the Boltzmann function 
may improve the estimate obtained by multiplying the CMB power spectrum 
by the suitable trigonometric functions of the rotation angle as in
Eqs.~(\ref{C_ll_EE_constant}-\ref{C_ll_TB_constant}). 

\section*{Acknowledgements} 
We wish to thank Daniela Paoletti, G{\"u}nter Sigl and Guido Zavattini for discussions.
FF and MG  are partially supported by INFN IS PD51
and by the ASI contract `Planck LFI Activity of Phase E2'.
FF is partially supported by INFN IS BO11.

\section*{Coulomb wave equation} 
\label{APP:1}
The Coulomb wave equation  is \cite{Abramowitz}:
\begin{equation}
\frac{d^2w}{dx^2}-\left[1-\frac{2 q}{x}-\frac{L(L+1)}{x^2}\right]w=0\,,
\end{equation}
with $x>0, -\infty<q<\infty, L$ a non negative integer.
Here, in order to solve Eq.~(\ref{eq:Coul}), we are particular interested to the particular case when L=0.

The solution can be written in terms of regular ($F_L(q,x)$) and irregular ($G_L(q,x)$) Coulomb wave function:
\begin{equation}
w=c_1 F_L(q,x) + c_2 G_L(q,x)\,.
\end{equation}

The Coulomb functions can be expanded for large values of $x$ \cite{Abramowitz}:
\begin{eqnarray}
F_0&=&g \cos \theta + f \sin \theta\,,\\
G_0&=&f\cos \theta-g \sin \theta\,,
\end{eqnarray}
similarly for the first derivative respect to $x$
\begin{eqnarray}
F_0^{\prime}&=&g^* \cos \theta + f^* \sin \theta\,,\\
G_0^{\prime}&=&f^*\cos \theta-g^* \sin \theta\,,
\end{eqnarray}
with $\theta\equiv x-q \ln 2 x+\arg\Gamma(1+ i q)$ and:
\begin{eqnarray*}
f=\sum_{k=0}^{\infty}f_k\,,&\quad& g=\sum_{k=0}^{\infty}g_k\,,\\
\quad f^*=\sum_{k=0}^{\infty}f^*_k\,,&\quad& g^*=\sum_{k=0}^{\infty}g^*_k\,,
\end{eqnarray*}
where:
\begin{eqnarray}
f_0=1\,,\quad f_{k+1}=a_k f_k-b_k g_k\,;\\
g_0=0\,,\quad g_{k+1}=a_k g_k+b_k f_k\,;\\
f_0^*=0\,,\quad f_{k+1}^*=a_k f_k^*-b_k g_k^*-\frac{f_{k+1}}{x}\,;\\
g_0^*=1-\frac{q}{x}\,,\quad g_{k+1}^*=a_k g_k^*+b_k f_k^*-\frac{g_{k+1}}{x}\,;\\
a_k=\frac{(2k+1)q}{2(k+1)x}\,,\quad b_{k}=\frac{q^2-k(k+1)}{2(k+1)x}\,.
\end{eqnarray}
Restricting to the first order:
\begin{eqnarray}
f=1+\frac{q}{2x}+\mathcal{O}\left(\frac{1}{x^2}\right)\,,\\
g=\frac{q^2}{2x}+\mathcal{O}\left(\frac{1}{x^2}\right)\,,\\
f^*=-\frac{q^2}{2x}+\mathcal{O}\left(\frac{1}{x^2}\right)\,,\\
g^*=1-\frac{q}{2x}+\mathcal{O}\left(\frac{1}{x^2}\right)\,.
\end{eqnarray}
Summarizing the asymptotic expansion of $F_L(q,x)$ and $F_L(q,x)$ for large values of $x$ is:
\begin{eqnarray}
F_0(q,x)\simeq \frac{q^2}{2x}\cos \theta+ \left(1+\frac{q}{2x} \right)\sin\theta\,,\\
G_0(q,x)\simeq \left(1+\frac{q}{2x} \right)\cos \theta-\frac{q^2}{2x}  \sin\theta\,,
\end{eqnarray}
and for the first derivative: 
\begin{eqnarray}
F_0^{\prime}(q,x)\simeq \left(1-\frac{q}{2x} \right)\cos \theta-\frac{q^2}{2x}  \sin\theta\,,\\
G_0^{\prime}(q,x)\simeq -\frac{q^2}{2x}\cos \theta- \left(1-\frac{q}{2x} \right)\sin\theta\,.
\end{eqnarray}


\begin{thebibliography}{99}

\bibitem{Peccei:1977hh}
  R.~D.~Peccei and H.~R.~Quinn,
  Phys.\ Rev.\ Lett.\  {\bf 38} (1977) 1440.

\bibitem{kimphysrept}
  J.~E.~Kim,
  Phys.\ Rept.\  {\bf 150} (1987) 1.

\bibitem{Kolb:1990vq}
  E.~W.~Kolb and M.~S.~Turner,
  \textit{``The Early universe,''}
  Front.\ Phys.\  {\bf 69} (1990) 1.

\bibitem{Raffelt:1996wa}
  G.~G.~Raffelt,
  \textit{``Stars As Laboratories For Fundamental Physics: The Astrophysics Of
  Neutrinos, Axions, And Other Weakly Interacting Particles,''}, 1996 Chicago, USA: Univ. Pr.

\bibitem{Sikivie:2006ni}
  P.~Sikivie,
  Lect.\ Notes Phys.\  {\bf 741}, 19 (2008)
  [arXiv:astro-ph/0610440].

\bibitem{Andriamonje:2007ew}
  S.~Andriamonje {\it et al.}  [CAST Collaboration],
  JCAP {\bf 0704} (2007) 010
  [arXiv:hep-ex/0702006].


\bibitem{Raffelt:2006cw}
  G.~G.~Raffelt,
  Lect.\ Notes Phys.\  {\bf 741}, 51 (2008)
  [arXiv:hep-ph/0611350].

\bibitem{Harari:1992ea}
  D.~Harari and P.~Sikivie,
  Phys.\ Lett.\  B {\bf 289} (1992) 67.

\bibitem{Carroll:1991zs}
  S.~M.~Carroll and G.~B.~Field,
  Phys.\ Rev.\  D {\bf 43} (1991) 3789.

\bibitem{Carroll:1997tc}
  S.~M.~Carroll and G.~B.~Field,
  Phys.\ Rev.\ Lett.\  {\bf 79} (1997) 2394
  [arXiv:astro-ph/9704263].

\bibitem{Lue:1998mq}
  A.~Lue, L.~M.~Wang and M.~Kamionkowski,
  Phys.\ Rev.\ Lett.\  {\bf 83} (1999) 1506
  [arXiv:astro-ph/9812088].

\bibitem{Zaldarriaga:1996xe}
  M.~Zaldarriaga and U.~Seljak,
  Phys.\ Rev.\ D {\bf 55} (1997) 1830
  [arXiv:astro-ph/9609170].

\bibitem{Zaldarriaga:1998rg}
  M.~Zaldarriaga,
  arXiv:astro-ph/9806122.

\bibitem{Hu:1997hv}
  W.~Hu and M.~J.~White,
  New Astron.\  {\bf 2} (1997) 323
  [arXiv:astro-ph/9706147].

\bibitem{Feng:2006dp}
  B.~Feng, M.~Li, J.~Q.~Xia, X.~Chen and X.~Zhang,
  Phys.\ Rev.\ Lett.\  {\bf 96} (2006) 221302
  [arXiv:astro-ph/0601095].

\bibitem{Cabella:2007br}
  P.~Cabella, P.~Natoli and J.~Silk,
  Phys.\ Rev.\  D {\bf 76}, 123014 (2007)
  [arXiv:0705.0810 [astro-ph]].

\bibitem{Komatsu:2008hk}
  E.~Komatsu {\it et al.}  [WMAP Collaboration],
  arXiv:0803.0547 [astro-ph].

\bibitem{Liu:2006uh}
  G.~C.~Liu, S.~Lee and K.~W.~Ng,
  Phys.\ Rev.\ Lett.\  {\bf 97}, 161303 (2006)
  [arXiv:astro-ph/0606248].
  
\bibitem{Balaji:2003sw}
  K.~R.~S.~Balaji, R.~H.~Brandenberger and D.~A.~Easson,
  JCAP {\bf 0312}, 008 (2003)
  [arXiv:hep-ph/0310368].
  
\bibitem{garretson}
  W.~D.~Garretson, G.~B.~Field and S.~M.~Carroll,
  Phys.\ Rev.\ D {\bf 46} (1992) 5346
  [arXiv:hep-ph/9209238].
  
\bibitem{finelli_1}
  F.~Finelli and A.~Gruppuso,
  Phys.\ Lett.\ B {\bf 502}, 216 (2001)
  [arXiv:hep-ph/0001231].

\bibitem{misner}
  C.~W.~Misner, K.~S.~Thorne and J.~A.~Wheeler, 
  \textit{``Gravitation,''},1973 San Francisco 

\bibitem{Kosowsky:1994cy}
  A.~Kosowsky,
  Annals Phys.\  {\bf 246} (1996) 49
  [arXiv:astro-ph/9501045].
  
\bibitem{Liu:2001xe}
  G.~C.~Liu, N.~Sugiyama, A.~J.~Benson, C.~G.~Lacey and A.~Nusser,
   Astrophys.\ J.\  {\bf 561}, 504 (2001)
  [arXiv:astro-ph/0101368].
  
\bibitem{Wu:2008qb}
  E.~Y.~Wu {\it et al.}  [QUaD Collaboration],
  arXiv:0811.0618 [astro-ph].

\bibitem{Dine:1982ah}
  M.~Dine and W.~Fischler,
  Phys.\ Lett.\  B {\bf 120}, 137 (1983).

\bibitem{Turner:1983he}
  M.~S.~Turner,
  Phys.\ Rev.\  D {\bf 28}, 1243 (1983).

\bibitem{footnote1}
Note that if we had included $k^2 ( |\tilde{A}_+^\prime |^2 -
|\tilde{A}_-^\prime |^2 )$ in the definition of $V$ in 
Eq.~(\ref{def:V}), the ratio in Eq.~(\ref{circulardegree}) would be proportional to 
${\cal O}(k^{-3})$.

\bibitem{Lee:1999ae}
  D.~S.~Lee and K.~W.~Ng,
  Phys.\ Rev.\  D {\bf 61}, 085003 (2000)
  [arXiv:hep-ph/9909282].


  
\bibitem{Gruppuso:2005xy}
  A.~Gruppuso and F.~Finelli,
  Phys.\ Rev.\  D {\bf 73}, 023512 (2006)
  [arXiv:astro-ph/0512641].
  
\bibitem{Lewis:1999bs}
  A.~Lewis, A.~Challinor and A.~Lasenby,
  Astrophys.\ J.\  {\bf 538}, 473 (2000)
  [arXiv:astro-ph/9911177].


\bibitem{Abbott:1982af}
  L.~F.~Abbott and P.~Sikivie,
  Phys.\ Lett.\  B {\bf 120}, 133 (1983).

\bibitem{Preskill:1982cy}
  J.~Preskill, M.~B.~Wise and F.~Wilczek,
  Phys.\ Lett.\  B {\bf 120}, 127 (1983).

\bibitem{Pi:1984pv}
  S.~Y.~Pi,
  Phys.\ Rev.\ Lett.\  {\bf 52}, 1725 (1984).

\bibitem{Linde:1987bx}
  A.~D.~Linde,
  Phys.\ Lett.\  B {\bf 201}, 437 (1988).

\bibitem{Tegmark:2005dy}
  M.~Tegmark, A.~Aguirre, M.~Rees and F.~Wilczek,
  Phys.\ Rev.\  D {\bf 73}, 023505 (2006)
  [arXiv:astro-ph/0511774].

\bibitem{Hertzberg:2008wr}
  M.~P.~Hertzberg, M.~Tegmark and F.~Wilczek,
  arXiv:0807.1726 [astro-ph].
  
\bibitem{Copeland:1997et}
  E.~J.~Copeland, A.~R.~Liddle and D.~Wands,
  Phys.\ Rev.\  D {\bf 57}, 4686 (1998)
  [arXiv:gr-qc/9711068].

\bibitem{Abramowitz}
   M. Abramowitz and I. A. Stegun,
 \textit{``Handbook of Mathematical Functions with Formulas, Graphs, and Mathematical Tables''}, 
 (Dover, New York, USA, 1964).

\bibitem{Anber:2006xt}
  M.~M.~Anber and L.~Sorbo,
  JCAP {\bf 0610} (2006) 018
  [arXiv:astro-ph/0606534].

\bibitem{Jain:2002vx}
  P.~Jain, S.~Panda and S.~Sarala,
  Phys.\ Rev.\  D {\bf 66} (2002) 085007
  [arXiv:hep-ph/0206046].

\bibitem{Mirizzi:2006zy}
  A.~Mirizzi, G.~G.~Raffelt and P.~D.~Serpico,
  Lect.\ Notes Phys.\  {\bf 741}, 115 (2008)
  [arXiv:astro-ph/0607415].


\bibitem{Das:2004qk}
  S.~Das, P.~Jain, J.~P.~Ralston and R.~Saha,
  JCAP {\bf 0506} (2005) 002
  [arXiv:hep-ph/0408198].

\end{thebibliography}
\end{document}